\documentclass[11pt]{article}
\usepackage{jheppub}

\usepackage{graphics}
\usepackage{amssymb}
\usepackage{amsfonts}
\usepackage{amsbsy}
\usepackage{amsmath}
\usepackage{slashed}
\usepackage[vcentermath]{youngtab}
\usepackage{graphicx}
\usepackage{amscd}
\usepackage{epstopdf}
\usepackage{amsthm}
\usepackage{enumerate}
\usepackage{pifont}
\usepackage{enumitem}
\usepackage{multirow}
\usepackage{booktabs}

\makeatletter
\newcommand{\mylabel}[2]{#2\def\@currentlabel{#2}\label{#1}}
\makeatother

\newcommand{\cC}{\mathcal{C}}

\newcommand{\cH}{\mathcal{H}}

\newcommand{\cM}{\mathcal{M}}
\newcommand{\cN}{\mathcal{N}}
\newcommand{\cO}{\mathcal{O}}

\newcommand{\cT}{\mathcal{T}}

\newcommand{\bC}{\mathbb{C}}

\newcommand{\bQ}{\mathbb{Q}}
\newcommand{\bR}{\mathbb{R}}

\newcommand{\bZ}{\mathbb{Z}}

\newcommand{\fg}{\mathfrak{g}}
\newcommand{\fh}{\mathfrak{h}}

\newcommand{\fS}{\mathfrak{S}}
\newcommand{\fu}{\mathfrak{u}}

\DeclareMathOperator{\tr}{tr}

\def\ov{\over}
\def\lam{{\lambda}}

\def\vev#1{\langle#1\rangle}

\def\eq#1{(\ref{#1})}

\def\Om{{\Omega}}

\def\a{{\alpha}}

\def \lam {\lambda}

\def \ra {\rightarrow}

\def\fg{{\mathfrak{g}}}
\def\fR{{\mathfrak{R}}}

\def\free{{\text{free}}}

\newcommand{\be}{\begin{equation}}
\newcommand{\ee}{\end{equation}}
\newcommand{\bea}{\begin{equation} \begin{aligned}}
\newcommand{\eea}{\end{aligned} \end{equation}}

\newcommand{\bln}{\begin{align}}
\newcommand{\eln}{\end{align}}
\newcommand{\bst}{\begin{split}}
\newcommand{\est}{\end{split}}
\newcommand{\bi}{\begin{itemize}}
\newcommand{\ei}{\end{itemize}}
\newcommand{\ben}{\begin{enumerate}}
\newcommand{\een}{\end{enumerate}}

\makeatletter
\def\saveenum{\xdef\@savedenum{\the\c@enumi\relax}}
\def\resetenum{\global\c@enumi\@savedenum}
\makeatother

\title{Quantization of anomaly coefficients in 6D $\mathcal{N}=(1,0)$ supergravity}

\author[\dagger]{Samuel Monnier,}
\author[\ddagger]{Gregory W. Moore}
\author[]{and}
\author[\ddagger]{Daniel S. Park}

\affiliation[\dagger]{
Section de Math\'ematiques, Universit\'e de Gen\`eve \\
2-4 rue du Li\`evre, 1211 Gen\`eve 4, Switzerland}
\affiliation[\ddagger]{
NHETC and Department of Physics and Astronomy\\
Rutgers University\\
Piscataway, NJ 08855, USA}

\emailAdd{{\tt samuel.monnier} {\rm at} {\tt gmail.com}}
\emailAdd{{\tt gmoore} {\rm at} {\tt physics.rutgers.edu}}
\emailAdd{{\tt d.s.park.hepth} {\rm at} {\tt gmail.com}}

\abstract{We obtain new constraints on the anomaly coefficients of 6D $\mathcal{N}=(1,0)$ supergravity theories using local and global anomaly cancellation conditions. We show how these constraints can be strengthened if we assume that the theory is well-defined on any spin space-time with an arbitrary gauge bundle. We distinguish the constraints depending on the gauge algebra only from those depending on the global structure of the gauge group. Our main constraint states that the coefficients of the anomaly polynomial for the gauge group $G$ should be an element of $2 H^4(BG;\mathbb{Z})\otimes \Lambda_S$ where $\Lambda_S$ is the unimodular string charge lattice. We show that the constraints in their strongest form are realized in F-theory compactifications. In the process, we identify the cocharacter lattice, which determines the global structure of the gauge group, within the homology lattice of the compactification manifold.}

\begin{document}

\setcounter{tocdepth}{2}

\maketitle

%
%

\section{Introduction}

A long-standing question is whether all consistent quantum gravity theories are string theories. While this question is not well posed in its most general form, it has led to fruitful discoveries with deep implications in quantum gravity and string theory when asked in the right context \cite{AlvarezGaume:1983ig, Green:1984sg, Gross:1985fr, Gross:1985rr, Seiberg:1988pf, Vafa:2005ui, Ooguri:2006in}. For example, the question of which ten dimensional $\cN=1$ supergravity theories are non-anomalous led to the first superstring revolution \cite{AlvarezGaume:1983ig, Green:1984sg, Gross:1985fr, Gross:1985rr}. It has been shown that the only allowed gauge algebras are given by $\mathfrak{so}(32)$ and $\mathfrak{e}_8 \oplus \mathfrak{e}_8$ \cite{AlvarezGaume:1983ig, Green:1984sg, Gross:1985fr, Gross:1985rr, Adams:2010zy}, both of which are  realized in string theory. String universality for six-dimensional supergravity theories with minimal supersymmetry \cite{StringUniversality,TaylorTASI}, the subject of this paper, has been studied along the same lines. While the subject has led to insights as to how string theories populate the so-called ``swampland" of supergravity theories that satisfy known consistency conditions, much remains to be understood, as both the string vacua and the supergravity landscape are richer in six dimensions.

The consistency conditions that have been studied in six-dimensional supergravity theories are local anomaly cancellation conditions \cite{AlvarezGaume:1983ig, Green:1984sg, Green:1984bx, Sagnotti:1992qw, Sadov:1996zm, Grassi:2000we}, the positivity of the gauge kinetic term \cite{Sagnotti:1992qw, Kumar:2009ae}, the absence of Witten anomalies \cite{Witten:1982fp, Bershadsky:1997sb, SuzukiTachikawa, KMT2}, and the unimodularity of the string charge lattice \cite{SeibergTaylor}. The space of supergravity theories that satisfy these consistency conditions is far bigger than the space of known string vacua \cite{KPT,TaylorTASI}. Conveniently, the latter can all be realized as F-theory compactifications \cite{Vafa:1996xn,Morrison:1996na,Morrison:1996pp} on elliptically fibered Calabi-Yau threefolds \cite{KMT1,TaylorTASI}.

Six-dimensional supergravity theories crucially involve a generalization of the Green-Schwarz mechanism for anomaly cancellation \cite{Green:1984sg, Green:1984bx, Sagnotti:1992qw, Sadov:1996zm}, and the corresponding couplings are called the \emph{anomaly coefficients}. The anomaly coefficients also determine the gauge invariant field strengths of the self-dual fields present in the tensor multiplets. In this paper, we use local and global \cite{Witten:1982fp, Witten:1985xe} anomaly cancellation, as well as the quantization of the string charges, to derive new constraints on the anomaly coefficients of six-dimensional supergravity theories. The string charge quantization condition is used along with the assumption that all spin spacetime manifolds and all gauge bundles should be included in the ``path integral" of the supergravity theory. This assumption can be seen as a strong version of the completeness hypothesis of \cite{Polchinski:2003bq, BanksSeiberg, Hellerman:2010fv}, which states that all charges allowed by Dirac quantization are realized in a consistent quantum gravity theory. We carefully state the assumptions required to derive each constraint, and distinguish the constraints depending only on the gauge Lie algebra from those depending on the global structure of the gauge group.

The strongest constraints on the gauge anomaly coefficients, which assume the strong version of the completeness hypothesis discussed above, can be formulated quite elegantly. The gauge anomaly coefficients can be assembled into a Weyl-invariant bilinear form $b$ on the Cartan subalgebra of the gauge Lie algebra, valued in the lattice $\Lambda_S$ of string charges. Such bilinear forms can be regarded as elements of
\begin{equation}
H^4(BG;\mathbb{R}) \otimes \Lambda_S ~ ,
\end{equation}
 where $BG$ is the classifying space of the gauge group $G$. Our main constraint then requires that 
\be
\frac{1}{2}b \in H^4(BG;\mathbb{Z}) \otimes \Lambda_S \, . 
\ee

We then show that all the newly derived constraints are satisfied in F-theory compactifications on elliptically fibered Calabi-Yau threefolds with smooth Calabi-Yau resolutions. Interestingly, the gravitational anomaly coefficient is always a characteristic vector of the string charge lattice in F-theory compactifications, something that does not follow from known constraints on the low energy theory (see the example in Section \ref{s: future}).

In the process, we identify a sublattice of the fourth homology lattice of the compactification manifold as the cocharacter lattice%
\footnote{See Appendix A for the definition of the various lattices involved. See also Appendix A of \cite{Gukov:2006jk} for a more thorough discussion.}
of the gauge group, thus identifying the global form of the gauge group of the F-theory compactification. Our approach to identifying the global structure of the gauge group provides a complementary approach to the existing literature, where the global structure is deduced from the algebraic properties of the elliptic fibration \cite{Aspinwall:1998xj}.%

As this work was being completed, we learned of the following related works. Certain constraints for 6D supergravity theories with a single abelian factor have been considered in \cite{Turner}. Constraint 5 of the present paper has been derived and used in \cite{Raghuram:2017qut}. Moreover, \cite{Garcia-Etxebarria:2017crf} investigates 8-dimensional compactifications of string theory along the same lines as the present work.

This paper is organized as follows. In Section \ref{s: review and summary} we review some basic facts about six-dimensional $\cN=(1,0)$ supergravity theories, including the known constraints on the anomaly coefficients. We then summarize the results of the paper. Section \ref{s: new} contains the derivation of the new constraints, and we show in Section \ref{s: Ftheory} that they always hold in F-theory. We also describe how to obtain the cocharacter lattice of the gauge group from the F-theory geometry there. We conclude in Section \ref{s: future}.

\section{Review and summary} \label{s: review and summary}

In this section, we first review the basics of six-dimensional $\cN=(1,0)$ supergravity theories. We also remind the reader of the known implications of anomaly cancellation. We then summarize the new constraints derived in the present paper.

\subsection{Six-dimensional $\cN=(1,0)$ supergravity theories}

\begin{table}[t!]
\center
 \begin{tabular}{@{}ll@{}}
 \toprule
 Multiplet & Field Content\\ \midrule
 Gravity & $(g_{\mu \nu}, \psi^+_\mu, B^+_{\mu \nu}) $  \\
 Tensor & $(B^-_{\mu \nu}, \chi^-, \phi)$   \\
 Vector & $(A_{\mu}, \lam^+)$   \\
 Hyper & $(\psi^-, 4\varphi)$  \\
 Half-hyper & $(\psi_\mathbb{R}^-, 2\varphi)$  \\ \bottomrule
  \end{tabular}
 \caption{The massless multiplets of 6D $\cN=(1,0)$ supergravity. The sign exponents indicate chirality. Whenever fermions are valued in the fundamental representation of the $SU(2)$ R-symmetry, a symplectic Majorana condition can be imposed. The gravity multiplet contains, in addition to the metric tensor $g_{\mu \nu}$, the gravitino $\psi^+_\mu$, which is a spin $3/2$ symplectic Majorana-Weyl fermion, and a real self-dual field $B^+_{\mu \nu}$. The tensor multiplet contain a real anti self-dual field $B^-_{\mu \nu}$, a symplectic Majorana-Weyl fermion $\chi^-$ and a real scalar $\phi$. The vector multiplet contains a gauge field $A_\mu$ and a symplectic Majorana-Weyl fermion $\lambda^+$. The hypermultiplet contains a Weyl fermion $\psi^-$ and a pair of complex bosons, yielding effectively four real bosonic degrees of freedom. The half-hypermultiplet can be constructed from a hypermultiplet valued in a quaternionic representation of the gauge group. In this case, a symplectic Majorana condition can be imposed on the fermions, and the bosons satisfy the corresponding reality condition. The half-hypermultiplet effectively contains a symplectic Majorana-Weyl fermion $\psi_\mathbb{R}^-$ and two real bosons.}
\label{t:mult}
\end{table}

Let us review the low-energy data characterizing six-dimensional $\cN=(1,0)$ supergravity theories.

\subsubsection{Massless spectrum}
The massless spectrum of the theory generally involves five types of multiplets, listed in Table \ref{t:mult}. It is summarized by the following data:
\ben
\item An integer $T \geq 0$, denoting the number of tensor multiplets in the theory, and an integral unimodular \cite{SeibergTaylor} lattice $\Lambda_S$ of string charges with signature $(1,T)$. The strings are here non-perturbative self-dual strings that are both electrically and magnetically charged under the self-dual 2-forms in the tensor multiplets. Given a choice of basis, we use the Greek letters $\alpha, \beta, \cdots$ to label the basis elements $\{e_\alpha\}$ of this lattice and denote the components of the nondegenerate bilinear form $\Omega$ on $\Lambda_S$ by $\Omega_{\alpha \beta} := \Omega(e_\alpha, e_\beta)$.
\item A compact reductive Lie group, the gauge group, whose Lie algebra is given by
\be\label{gauge}
\mathfrak{g}
= \mathfrak{g}_{\rm ss} \oplus \mathfrak{g}_{\rm a}
= \bigoplus_i \mathfrak{g}_i \oplus \bigoplus_I \mathfrak{u}(1)_I \;.
\ee
$\mathfrak{g}_{\rm ss} := [\fg, \fg]$ is the semi-simple part  of the gauge algebra and $\mathfrak{g}_{\rm a}$ is the abelian part. $i = 1,...,s$ labels the simple summands of $\mathfrak{g}_{\rm ss}$ and $I = 1,...,r$ labels the one-dimensional summands of $\mathfrak{g}_{\rm a}$. On the simple summands, we normalize the Killing form so that the long roots have norm square 2. In our conventions, the exponential map is $x \rightarrow \exp(ix)$, $x \in \mathfrak{g}$, so that the Lie algebra elements are hermitian matrices in unitary representations.

The component of the gauge group connected to the identity has the form
\be
\label{EqGlobFormGaugeGroup}
G = (\tilde{G}_{\rm ss} \times G_{\rm a})/\Gamma \;,
\ee
where $\tilde{G}_{\rm ss}$ is a semi-simple simply connected compact group with Lie algebra $\mathfrak{g}_{\rm ss}$ and $G_{\rm a} \simeq U(1)^r$ is an abelian group. If $Z$ denotes the center of $\tilde{G}_{\rm ss}$, $\Gamma$ is a subgroup of $Z \times G_{\rm a}$ intersecting $1 \times G_{\rm a}$ trivially. We will not consider possible constraints associated to the components of the gauge group disconnected from the identity, so in the following, we will simply refer to $G$ as ``the gauge group." $\tilde{G}_{\rm ss} = \prod_i \tilde{G}_i$ is a direct product of simple simply connected compact Lie groups $\tilde{G}_i$ with ${\rm Lie}(\tilde{G}_i) = \mathfrak{g}_i$. We also write $\tilde{G} := \tilde{G}_{\rm ss} \times G_{\rm a}$. Note that this is not the simply connected cover of $G$, as the latter is non-compact in general.
\footnote{The groups $G$ and $\tilde G$ are, in general, not semisimple but rather reductive. 
We will nevertheless use some terminology usually reserved for semisimple groups. We define the maximal torus 
of $\tilde G$ and $G$ to be $\tilde T := \tilde T_{\rm ss} \times G_{\rm a}$ and $T:=\tilde T/\Gamma$, 
respectively. The Cartan subalgebras are the tangent space at the identity of these tori. The coroot 
lattice is defined to be $\Lambda^{G}_{\rm CR} := {\rm Hom}(U(1), \tilde T)$ while the cocharacter 
lattice is $\Lambda^{G}_{\rm CC} := {\rm Hom}(U(1),T)$. Note that the coroot lattice is the cocharacter 
lattice of $\tilde G$. Dually, the character lattice of $G$ is $\Lambda^G_{\rm C} := {\rm Hom}(T,U(1))$. 
The weight lattice is the character lattice of $\tilde G$.}

\item The matter lies in a representation $\mathfrak{R}$ of $G$. The representation is completely reducible
and can be decomposed as:
\be\label{hypers}
\fR = \bigoplus_a \fR^a = \bigoplus_a ~\left( \bigotimes_i \fR^a_i \otimes \bigotimes_I q^a_I \right) \;.
\ee
where the  $\fR^a$'s are irreducible representations of $G$. Different values of $a$ can give isomorphic representations.  The
 $\fR^a$ are tensor products of irreducible representations of the summands $\mathfrak{g}_i$ and $\mathfrak{u}(1)_I$. $q^a_I \in \mathbb{Z}$ are the charges of the representation under $\mathfrak{u}(1)_I$. The abelian charges are normalized so that the smallest allowed charge is $1$. We assume that the $\fR^a$'s are complex, and therefore that there are no half-hypermultiplets. Some of the constraints we will derive are independent of the matter content, and therefore also valid in the presence of half-hypermultiplets. We will indicate when this is the case.

\een

Given this data, one can compute the degree 8 anomaly polynomial $I_8$ of the theory by adding up the contributions of all the chiral fields \cite{AlvarezGaume:1983ig}, i.e. the gravitino, self-dual/anti-self-dual tensor fields, and chiral fermions.

While there can be strongly coupled SCFTs \cite{Witten:1995ex, Witten:1995zh, WittenSmall,
Strominger:1995ac, Ganor:1996mu, SW6, Bershadsky:1996nu, Blum:1997fw, instK3, Intriligator:1997dh, Brunner:1997gf, Hanany:1997gh, HMV, DHTV, Heckman, Heckman:2015bfa} in the effective six-dimensional supergravity theory \cite{DelZotto:2014fia, Anderson:2015cqy}, we do not consider that possibility in this work. We note that all non-trivial six-dimensional SCFTs with $\cN = (1,0)$ supersymmetry constructed in F-theory have a tensor branch, whose effective theory can be described in terms of the massless multiplets reviewed in this section \cite{HMV, DHTV, Heckman, Heckman:2015bfa}.

\subsubsection{Anomaly coefficients}

\label{SecAnCoeff}

The invariant field strengths for the tensor fields $B^\alpha$ in the theory are given by
\be
H^\alpha = dB^\alpha + \Gamma^\alpha
\ee
where $\Gamma^\alpha$ is a Chern-Simons three-form associated to the 4-form:
\be
\label{EqDefCSGamma}
d \Gamma^\alpha = Y^\alpha = \frac{a^\alpha}{16 \pi^2}  \tr R^2 + {\sum_i} \frac{b_i^\alpha}{8 \pi^2} \tr F_i^2
+  \sum_{I,J} \frac{b^\alpha_{IJ}}{8 \pi^2} F_I F_J \,,
\ee
where $\tr$ denotes the trace in the adjoint representation divided by twice the dual Coxeter number, i.e. normalized so that the long roots have length square 2.

$\Gamma^\alpha$ can be interpreted as an effective degree 3 abelian gauge field with field strength $Y^\alpha$, coupled to the tensor multiplets through the Green-Schwarz terms (see below). In order to investigate the flux quantization of this effective gauge field, it is useful to reexpress \eqref{EqDefCSGamma} in terms of characteristic forms:
\be
Y^{\alpha} =   \frac{1}{4} a^\a p_1 - \sum_i b^\a_i c_2^i + \frac{1}{2}\sum_{IJ} b^\a_{IJ} c_1^I c_1^J \,.
\ee
$p_1 := \frac{1}{8\pi^2} \tr_{\rm vec} R^2 = \frac{1}{4\pi^2} \tr R^2$ is the first Pontryagin form of the tangent bundle, with $\tr_{\rm vec}$ the trace in the vector representation.  $c_1^I := F_I/2\pi$ is the first Chern form of the $I$th $U(1)$ bundle. $c_2^i := -\frac{1}{8 \pi^2} \tr F_i^2$ is the ``second Chern form" associated to $\mathfrak{g}_i \subset \mathfrak{g}$, namely a degree 4 characteristic form associated to the quadratic Casimir of $\mathfrak{g}_i$. $c_2^i$ coincides with the standard second Chern form of a complex vector bundle
with connection in the case $\mathfrak{g}_i = \mathfrak{su}(n)$. We discuss the properties of $c_2^i$ further in Appendix \ref{AppIntc2}, in particular the quantization of its periods in relation to the global structure of the gauge group.

The vectors
\be
a,~b_i,~b_{IJ}=b_{JI} \quad \in \quad \Lambda_S \otimes \mathbb{R}
\ee
are called the {\it anomaly coefficients} of the theory. $a$ is the gravitational anomaly coefficient, while $b_i$ and $b_{IJ}$ are the non-abelian and abelian anomaly coefficients, respectively.

Note that $Y$ is a degree four form and therefore must be quadratic in the curvatures $R$ and $\{F_i, F_I\}$. Since two-form commute, the component of $Y$ quadratic in the gauge curvatures is a symmetric gauge invariant bilinear form on $\mathfrak{g}$. We can therefore see $b_i$ and $b_{IJ}$ as parametrizing a $\Lambda_S \otimes \mathbb{R}$-valued Weyl-invariant bilinear form on the Cartan subalgebra of $\mathfrak{g}$. Let $K_i$ be the canonically normalized Killing form on $\mathfrak{g}_i$, with respect to which the long roots have length 2. Then the bilinear form associated to $\{b_i, b_{IJ}\}$ is
\be
\label{EqBilinFormAnomCoeff}
b = b_{\rm ss} \oplus b_{\rm a} := \bigoplus_i b_i K_i \oplus (b_{IJ}) \;,
\ee
where $(b_{IJ})$ is the bilinear form on $\mathbb{R}^r \simeq \mathfrak{u}(1)^r$ having matrix elements $b_{IJ}$. The decomposition of $b$ into components $\{b_i, b_{IJ}\}$ has some arbitrariness in the sense that the latter are acted upon by automorphisms of $G$. The invariant object capturing the gauge anomaly coefficients is the bilinear form $b$. We will see in Section \ref{s: Ftheory} that $b$ has a concrete geometric realization in F-theory.

\subsubsection{Self-duality condition and the modulus}

\label{SecSDCondMod}

In order to define the self-dual and anti-self dual conditions, a vector $j \in \Lambda_S \otimes \bR$ must be introduced, with unit norm with respect to $\Omega$. This ``modulus" parametrizes the vacuum expectation values of the $T$ scalar fields in the tensor multiplets. We now introduce the notation
\be
\Om^{\alpha \beta} = (\Om^{-1})_{\alpha \beta} \,,
\ee
and agree to raise and lower $\alpha$ indices with $\Om$. We then define
\be
G_{\alpha \beta} = 2j_\alpha j_\beta - \Om_{\alpha\beta} \,,
\ee
where $j_\alpha = \Om_{\alpha \beta} j^\beta$ are the components of the vector dual to $j$. The self-duality conditions are then given by
\be
* \Om_{\alpha \beta} H^\beta = G_{\alpha \beta} H^{\beta} \,.
\ee
Note that this implies that
\be
* j \cdot H = j \cdot H
\ee
while for any $e \in \Lambda_S$ with $e \cdot j = 0$,
\be
* e \cdot H = - e \cdot H \,.
\ee

The magnetic charge source $\widetilde{J}$ is given by
\be
\widetilde{J}^\alpha = dH^\alpha = Y^\alpha \,,
\ee
the electric source is given by
\be
J_\alpha = d * G_{\alpha \beta}H^\beta = dH_\alpha
= \Om_{\alpha \beta} \widetilde{J}^\beta \,.
\ee
Note that $J$ defines an element of cohomology
 $[J ] \in \Lambda_S \otimes H^4(M;\mathbb{R})$, while $[\tilde{J}] \in \Lambda_S^\ast \otimes H^4(M;\mathbb{R})$, where $M$ is the spacetime manifold. Denoting by $()^\vee$ the identification of $\Lambda_S \otimes H^4(M;\mathbb{R})$ with $\Lambda_S^\ast \otimes H^4(M;\mathbb{R})$ induced by $\Omega$, we have:
\be
[J] =[ \widetilde{J}]^\vee \,,
\ee
i.e. the electric source is the dual of the magnetic source, as it should be.

The factor
\be\label{GS}
\exp \left(- 2 \pi i \cdot \frac{1}{2} \int B^\alpha \wedge J_\alpha
\right)
= \exp \left(- 2 \pi i \cdot \frac{1}{2} \int \Om_{\alpha\beta}
B^\alpha \wedge Y^\beta
\right)
\ee
is included in the ``path integral" of the theory. \eqref{GS} is the coupling between the tensor fields and their electric sources \cite{SW}.

Let us remark here that there are serious problems with the expression \eqref{GS}, which will be solved in an upcoming publication. As written, \eqref{GS} assumes that the gauge fields $B^\alpha$ are representable by 2-forms, and are therefore topologically trivial. \eqref{GS} looks like the boundary value of a standard 7-dimensional Chern-Simons coupling of the form $\Gamma^\alpha d \Gamma^\alpha$, so one may try to define the latter on generic fields in the standard way, using the integral of characteristic forms on a 8-dimensional manifold bounded by spacetime, or alternatively using differential cocycles. There is, however, a factor $\frac{1}{2}$ in \eqref{GS} that makes such definitions gauge invariant only up to a sign under large gauge transformations. It turns out that the Chern-Simons term generating the Green-Schwarz terms \eqref{GS} is an exponentiated Wu Chern-Simons action \cite{Monnier:2016jlo}, obtained from a Lagrangian valued in a generalized cohomology theory. This is completely parallel to the appearance of generalized cohomology theories in the Wess-Zumino terms of the non-linear sigma model describing pions in QCD \cite{Freed:2006mx}. A similar factor of one-half also appears in the holographic 11-dimensional Chern-Simons action for the type IIB self-dual fields \cite{Belov:2006xj}.

\subsubsection{The generalized completeness hypothesis}

 We describe here a plausible conjecture about supergravity. Recall that the {\it completeness hypothesis} is the conjecture that all the charges allowed by the Dirac quantization condition are present in consistent quantum gravity theories. Arguments supporting it have appeared in \cite{Polchinski:2003bq, BanksSeiberg, Hellerman:2010fv}. This conjecture is not necessarily true for field theories, in which one may choose to sum over a subset of distinct topological sectors in the path integral \cite{Pantev:2005rh, Pantev:2005zs, SeibergTop, BanksSeiberg}.

We will use in the present paper a stronger version of the completeness hypothesis, stating in addition that a consistent supergravity theory may be put on an arbitrary spin manifold, and that any smooth gauge field configuration should be allowed in the supergravity ``path integral". This generalizes the completeness hypothesis for the following reason. The Green-Schwarz mechanism involves a trivialization of the degree 4 characteristic form $Y$. On a general spacetime endowed with a general gauge bundle, $Y$ is non-trivial and carries background charges. Those charges have to be canceled by introducing background strings. The completeness hypothesis ensures that these strings exist in the theory; it is therefore required before starting to consider generic backgrounds.
\footnote{It would be interesting to investigate the possibility of using symplectic Majorana-Weyl conditions to generalize
this completeness hypothesis yet further to include ${\rm Spin}^c$, but not ${\rm Spin}$ manifolds. But that goes well beyond
the scope of this paper.}

This conjecture allows us to  choose freely the spacetime and the gauge bundle when examining the consistency of six-dimensional supergravity theories. Let us however emphasize that we do not assume it a priori. We will explicitly state which constraints rely on it.

\subsubsection{Anomaly cancellation conditions}

The term in the exponential of equation \eq{GS} is the Green-Schwarz term of the supergravity theory. In order for the local gravitational, gauge and mixed anomalies of the theory to be cancelled through the Green-Schwarz mechanism \cite{Green:1984sg}, the Green-Schwarz-Sagnotti-West anomaly cancellation condition \cite{Green:1984bx, Sagnotti:1992qw, Sadov:1996zm} must be satisfied. This condition requires that the anomaly polynomial $I_8$, determined by the massless spectrum of the theory, factorizes as follows:
\be\label{factorization}
\frac{1}{2 \pi i}I_8 = \frac{1}{2} (Y,Y) = \frac{1}{2} \Om_{\alpha \beta} Y^\alpha \wedge Y^\beta \,.
\ee
The local anomaly cancellation conditions can be written by comparing the left and right hand side of equation \eq{factorization} term by term. We will first state them, and then explain the notation involved. The cancellation conditions that do not involve the abelian anomaly coefficients are given by
\bea\label{GSSW}
0&=H-V+29T-273\,,&
a \cdot a &= 9-T \,, \\
a \cdot b_i &=
{1 \ov 6} \left(A_{\text{Adj}_i}
- \sum_{\fR_i} x^i_{\fR_i} A_{\fR_i}\right) \,, &
0&= B_{\text{Adj}_i}
- \sum_{\fR_i} x^i_{\fR_i} B_{\fR_i} \,, \\
b_i \cdot b_i &=
{1 \ov 3} \left(-C_{\text{Adj}_i}
+ \sum_{\fR_i} x^i_{\fR_i} C_{\fR_i}\right) \,, &
b_i \cdot b_j
&= \sum_{\fR_i, \fS_j} x^{i,j}_{\fR_i, \fS_j} A_{\fR_i} A_{\fS_j} ~(i \neq j)\,,
\eea
while those that do are given by
\bea
\label{GSSW abelian}
a \cdot b_{IJ} &=
-{1 \ov 6}
\sum_{\fR_i} x^{I,J}_{q_I, q_J} q_I q_J \\
0&= \sum_{\fR_i, q_I} x^{i,I}_{\fR_i,q_I}  q_I  E_{\fR_i}\,, \\
b_i \cdot b_{IJ}
&= \sum_{\fR_i q_I,q_J} x^{i,I,J}_{\fR_i,q_I,q_J} q_I q_J  A_{\fR_i}\,, \\
b_{IJ} \cdot b_{KL}
+b_{IL} \cdot b_{KJ}
+b_{IK} \cdot b_{JL} &= \sum_{q_I,q_J,q_K,q_L} x^{I,J,K,L}_{q_I,q_J,q_L,q_K} q_I q_J q_L q_K \,.
\eea
$H$ is the total number of hypermultiplets in the theory, while $V$ is the total number of vector multiplets:
\be
H = {\rm dim}_{\mathbb{C}} \, \fR\,, \qquad
V = {\rm dim}_{\mathbb{R}} \, \mathfrak{g} \,.
\ee
The Lie algebra coefficients $A$, $B$, $C$ and $E$ are defined by
\bea\label{ABCE}
\tr_{\fR_i} F^2 &= A_{\fR_i} \tr F^2 \,, &
\tr_{\fR_i} F^4 &= B_{\fR_i} \tr F^4 +
C_{\fR_i} (\tr F^2)^2  \,, &
\tr_{\fR_i} F^3 &= E_{\fR_i} \tr F^3
\eea
for $F \in \mathfrak{g}_i$. For $\mathfrak{g}_i$ with rank $\leq 2$, $B_{\fR_i}$ is defined to vanish for any $\fR_i$. Informally, $x^{i_1,\cdots,i_k, I_1,\cdots,I_K}_{\fR_{i_1}, \cdots,\fR_{i_k},  q_{I_1}, \cdots, q_{I_K}}$ is defined to be the number of hypermultiplets simultaneously in representation $\fR_{i_1}$ of $\fg_{i_1}$, ..., $\fR_{i_k}$ of $\fg_{i_k}$, $q_{I_1}$ of $\fu(1)_{I_1}$, ... and $q_{I_K}$ of $\fu(1)_{I_K}$, where the representations with respect to the unspecified summands can be arbitrary. More precisely,
\be
x^{i_1,\cdots,i_k, I_1,\cdots,I_K}_{\fR_{i_1}, \cdots,\fR_{i_k},
q_{I_1}, \cdots, q_{I_K}}
=  \sum_{\substack{a~:~\fR^a_{i_1} = \fR_{i_1},\cdots, \fR^a_{i_k} = \fR_{i_k} \\
q^a_{I_1}=q_{I_1},\cdots,q^a_{I_K} =q_{I_K} }}
\left( \prod_{i \notin \{ i_1,\cdots,i_k\}} { {\rm dim}_\mathbb{C}  \fR^a_i} \right)
\ee
$a$ labels the irreducible representations of $\fR$ as in equation \eq{hypers}. The $x$'s are obviously always integers. For the rest of the paper, we use the abbreviated notation $x_{\fR_{i_1}, \cdots,\fR_{i_k},  q_{I_1}, \cdots, q_{I_K}}$ which is understood to be equal to $x^{i_1,\cdots,i_k, I_1,\cdots,I_K}_{\fR_{i_1}, \cdots,\fR_{i_k},  q_{I_1}, \cdots, q_{I_K}}$ as defined above. The upper case indices of equation \eq{GSSW abelian} may be degenerate, and we use the convention
\be
x_{q_I,q_I,q_I,q_I} = x_{q_I} \,, \quad
x_{q_I,q_I,q_I,q_J} = x_{q_I,q_J} \,, \quad ...
\ee
Thus, for example, the last equation of \eq{GSSW abelian} implies the following five equations when $I$, $J$, $K$ and $L$ are assumed to be distinct:
\bea
3 b_{II} \cdot b_{II} &= \sum_{q_I} x_{q_I} q_I^4 \,, \\
3 b_{II} \cdot b_{IJ} &= \sum_{q_I,q_J} x_{q_I} q_I^3 q_J \,, \\
b_{II} \cdot b_{JJ} + 2 b_{IJ} \cdot b_{IJ}
&= \sum_{q_I,q_J} x_{q_I} q_I^2 q_J^2 \\
b_{II} \cdot b_{JK} +  2b_{IJ} \cdot b_{IK}
&= \sum_{q_I,q_J,q_K} x_{q_I,q_J,q_K} q_I^2 q_J q_K \\
b_{IJ} \cdot b_{KL}
+b_{IL} \cdot b_{KJ}
+b_{IK} \cdot b_{JL} &= \sum_{q_I,q_J,q_K,q_L} x_{q_I,q_J,q_L,q_K} q_I q_J q_L q_K \,.
\eea
\\
\paragraph{Remark} Supersymmetry implies that the coefficients of the gauge kinetic terms of the non-abelian and abelian gauge fields are positive multiples of
\be
j \cdot b_i \,, \quad {j \cdot b_{IJ}} \,,
\ee
respectively. Thus unitarity implies that there must exist a unit vector $j \in \Lambda_S \otimes \bR$ such that $j \cdot b_i$ is positive for all $i$ and $j \cdot b_{IJ}$ is a positive definite matrix.

Finding massless spectra for which $a$, $b_i$ and $b_{IJ}$ that satisfy the unitarity constraint above and solve the anomaly equations is a non-trivial task. For example, it has been shown that only a finite number of gauge algebras have a unitary solution to the anomaly cancellation conditions when $T<9$, and furthermore, that only a finite number of gauge algebra-matter representation combinations have unitary solution when the gauge algebra is assumed to be non-abelian \cite{Kumar:2009ae,KPT,PT}.

\paragraph{The string charge quantization condition}  The Green-Schwarz term \eqref{GS} implies that $Y$ carries string charge. On a compact manifold, the string charge, like any charge, has to vanish globally. As $Y$ is not necessarily topologically trivial, the corresponding charge has to be canceled by background self-dual strings. Obviously, this is only possible when the integral of $Y$ along any integral 4-cycle $\Sigma_4$ yields an element of the string charge lattice $\Lambda_S$. Explicitly,
\be\label{scq 2}
\int_{\Sigma_4} Y = a \int_{\Sigma_4} {1 \ov 4}p_1
-\sum_i b_i \int_{\Sigma_4} c_2^i
+{1 \ov 2} \sum_{IJ} b_{IJ}  \int_{\Sigma_4} c_1^I c_1^J \in \Lambda_S \;,
\ee
where $M$ is the spacetime manifold. We will refer to \eqref{scq 2} as the "string charge quantization condition". If $\int_{\Sigma_4} Y = x \in \Lambda_S$, with $x \neq 0$, then a self-dual string of charge $x$ has to be wrapped along the Poincar\'e dual 2-cycle in $M$. The constraints imposed by \eqref{scq 2} on the anomaly coefficients depend on the possible periods of $p_1$, $c_2^i$ and $c_1^I$. The generalized completeness hypothesis allows us to obtain strong constraints by evaluating \eqref{scq 2} on suitably chosen spacetimes and gauge bundles.

\subsection{Known constraints on anomaly coefficients}

A consequence of the anomaly cancellation conditions is that the inner-products between the the vectors $a$ and $b_i$ are integral \cite{KMT2}:
\be
a\cdot b_i\,,~b_i \cdot b_j~\in~\bZ\quad
\text{for all }i,j \,.
\ee
These constraints follow from both local and global anomaly cancellation. In particular, given a gauge algebra $\mathfrak{g}_i$ with rank $\leq 2$, the sixth homotopy group $\pi_6$ of its simply connected gauge group $\widetilde{G}_i$ has a non-trivial generator $g$. There is a canonical map $f:\widetilde{G}_i \rightarrow G$ from $\widetilde{G}_i$ to the global form of the gauge group $G$ of the supergravity theory. The phase of the global gauge transformation on $S^6$ by $f(g)$ can be computed by embedding $\widetilde{G}_i$ into a gauge group of higher rank \cite{ElitzurNair,Bershadsky:1997sb,SuzukiTachikawa,KPT}. The integrality of $a \cdot b_i$ and $b_i^2$ follows from assuming that this phase vanishes.

It has been shown \cite{SeibergTaylor} that the consistency of the theory upon reduction to lower dimensions requires that:
\be
b_i ~\in~ \Lambda_S \,.
\ee

\subsection{New constraints and summary of results}

\label{SecNewConstrRes}

In this paper, we find the following constraints on the anomaly coefficients:
\begin{enumerate}
\item \label{Constr_a_char}$a \cdot b_i + b_i \cdot b_i \in 2 \bZ$.
\item \label{Constr_2}$6 a \cdot b_{IJ} \in \bZ$.
\item \label{Constr_3}$b_i \cdot b_{IJ} \in \bZ$.
\item \label{Constr_4}$(b_{IJ} \cdot b_{KL} + b_{IK} \cdot b_{JL} + b_{IL} \cdot b_{JK}) \in \bZ$ for any $I,J,K,L$.
\item \label{Constr_5}$b_{II}^2 +2 a \cdot b_{II}  \in 4 \bZ$.
\item \label{Constr_6}$b_{II}\cdot b_{IJ} +2 a \cdot b_{IJ} \in 2 \bZ$.
\item \label{Constr_7}$b_{II}\cdot b_{JJ} + 2b_{IJ}\cdot b_{IJ} +6 a\cdot b_{IJ} \in 2 \bZ$.
\item \label{Constr_8}$b_i \cdot b_{II} \in 2 \bZ$ for $\mathfrak{g}_i =\mathfrak{su(n \geq 3)}, ~\mathfrak{so(n \geq 6)}, ~\mathfrak{g_2}, ~\mathfrak{f_4}, ~\mathfrak{e_{6,7,8}}$.
\end{enumerate}
The constraints above are all derived using the equations \eqref{GSSW} and \eqref{GSSW abelian} satisfied by the anomaly coefficients, and in particular do not rely on the generalized completeness hypothesis.

If we do assume the generalized completeness hypothesis
 we can use global anomalies to derive the analogue of Constraint \ref{Constr_8} for $\mathfrak{g}_i = \mathfrak{sp}(n)$ (with $\mathfrak{su}(2) \simeq \mathfrak{sp}(1)$):
\saveenum
\begin{enumerate}
\resetenum
\item[8*.]$b_i \cdot b_{II} \in 2 \bZ$ for $\mathfrak{g}_i = \mathfrak{sp}(n)$.
\end{enumerate}
Using the generalized completeness hypothesis in conjunction with the string charge quantization condition \eqref{scq 2}, we will derive
\saveenum
\begin{enumerate}
\resetenum
\item \label{Constr_10}$a,~b_i,~{1 \ov 2} b_{II},~b_{IJ}~ \in  ~\Lambda_S$.
\end{enumerate}

A more conceptual way of interpreting constraint \ref{Constr_10}  is the following. Recall from \eqref{EqBilinFormAnomCoeff} that the anomaly coefficients can be gathered into a $\Lambda_S \otimes \mathbb{R}$-valued Weyl-invariant bilinear form $b = b_{\rm ss} \oplus b_{\rm a}$ on the Cartan subalgebra of $\mathfrak{g}$. The coroots are the Cartan elements that are sent to the identity of the group by the exponential map of $\tilde{G}$. Let $\Lambda_{\rm CR}$ be the lattice of coroots inside the Cartan subalgebra. We can restrict $b$ to a bilinear form on $\Lambda_{\rm CR}$. Constraint \ref{Constr_10} says that $b$ is an \emph{even $\Lambda_S$-valued} Weyl invariant bilinear form on $\Lambda_{\rm CR}$, in the sense that
\be
b(x, y)\in \Lambda_S \;, \quad b(x, x)\in 2 \Lambda_S
\ee
for any coroots $x,y$. Indeed, $b_{\rm ss}$ is even $\Lambda_S$-valued on $\Lambda_{\rm CR}$ whenever $b_i \in \Lambda_S$, as the canonically normalized Killing forms $K_i$ are even on $\Lambda_{\rm CR}$. Moreover, as the coroots of $\mathfrak{g}_{\rm a} \simeq \mathbb{R}^r$ are just the vectors with integer components, Constraint \ref{Constr_10} explicitly forces $b_{\rm a}$ to be even $\Lambda_S$-valued. $b$ being even $\Lambda_S$-valued on $\Lambda_{\rm CR}$ is equivalent to $\frac{1}{2}b(x,x)$ being an element of $\Lambda_S$ for every coroot $x$.

Now (see also Appendix \ref{AppIntc2}), the space of Weyl invariant bilinear forms on the Cartan subalgebra of $\mathfrak{g}$ coincides with $H^4(BG;\mathbb{R}) \simeq \mathbb{R}^{s+r(r+1)/2}$. $H^4(B\tilde{G};\mathbb{Z})$ and $H^4(BG;\mathbb{Z})$, being free groups isomorphic to $\mathbb{Z}^{s+r(r+1)/2}$ \cite{2016arXiv160202968H}, are naturally embedded into $H^4(BG;\mathbb{R})$. Hence they can be interpreted as invariant bilinear forms satisfying certain integrality conditions: $H^4(B\tilde{G};\mathbb{Z})$ and $H^4(BG;\mathbb{Z})$ are the bilinear forms giving integer norm square to the coroots and cocharacters of $G$, respectively. Constraint \ref{Constr_10} is then equivalent to
\saveenum
\begin{enumerate}
\resetenum
\item[9*.] \label{Constr_10'}$a \in \Lambda_S$, $\frac{1}{2}b \in H^4(B\tilde{G};\mathbb{Z}) \otimes \Lambda_S$.
\end{enumerate}
It turns out that $H^4(BG;\mathbb{Z}) \subset H^4(B\tilde{G};\mathbb{Z})$, and that when taking into account the global structure of the gauge group, Constraint \ref{Constr_10} is strengthened to
\saveenum
\begin{enumerate}
\resetenum
\item \label{Constr_11}$a \in \Lambda_S$, $\frac{1}{2}b \in H^4(BG;\mathbb{Z}) \otimes \Lambda_S$.
\end{enumerate}
Constraint \ref{Constr_10} and Constraint \ref{Constr_11} are equivalent to $b$ being an even $\Lambda_S$-valued bilinear form when restricted to the coroot and cocharacter lattice of $G$, respectively.

\paragraph{Remarks}
\begin{itemize}

\item The proof of Constraint 8* is redundant, as Constraint 8* follows directly from Constraint \ref{Constr_10} with the same assumptions. We find it nevertheless instructive enough to include it, as it involves global anomalies.

\item Constraints \ref{Constr_a_char}-\ref{Constr_8} above follow from Constraint \ref{Constr_10} if $a$ is a characteristic element of $\Lambda_S$, i.e. if $a \cdot x + x \cdot x \in 2\mathbb{Z}$ for all $x \in \Lambda_S$. Although we could not derive this extra constraint in the context studied in the present paper, it is satisfied in F-theory compactifications and follows from the careful construction of the Green-Schwarz terms hinted at in Section \ref{SecSDCondMod}.

\item One may wonder about a more systematic approach to the computation of global anomalies. The latter are generally computed by evaluating eta invariants of 7-dimensional mapping tori. When a 7-dimensional mapping torus bounds an 8-manifold over which the gauge bundle extends, it can be reexpressed in terms of the local index density on the 8-manifold, so the cancellation of global anomalies is equivalent to the cancellation of local anomalies. When there is no such 8-manifold, i.e. when the 7-dimensional mapping torus belongs to a non-trivial bordism class, global anomalies can exist in the absence of local anomalies. Unfortunately, eta invariants are extremely difficult to compute, which is why a systematic computation of global anomalies is not possible. A smarter approach would be to identify and characterize the bordism invariant corresponding to the global anomaly. A good candidate for the relevant anomaly field theory is a certain Chern-Simons-Wu theory which will be described elsewhere.

\item In Section \ref{s: Ftheory}, we explore the constraints satisfied by the anomaly coefficients of supergravity theories obtained from F-theory compactifications. We find that Constraint \ref{Constr_11} holds, and that $a$ is a characteristic element of $\Lambda_S$. By the discussion above, we conclude that Constraints \ref{Constr_a_char}-\ref{Constr_11} automatically hold in F-theory compactifications.

\item While we could derive Constraints \ref{Constr_a_char}-\ref{Constr_8} only in theories without half-hypermultiplets, Constraints \ref{Constr_10} and \ref{Constr_11} are valid in the presence of half-hypermultiplets. Their derivation does not refer at all to the matter content of the theory (see Section \ref{ss: charge quantization}). Concerning the possible validity of Constraints \ref{Constr_a_char}-\ref{Constr_8} in the presence of half-hypermultiplets, let us remark that the F-theory analysis does not make any assumptions about non-existence of half-hypermultiplets, and we show that the anomaly coefficients do satisfy Constraints \ref{Constr_a_char}-\ref{Constr_8} in F-theory. Many vacua with half-hypermultiplets can be realized in F-theory \cite{Grassi:2000we, Morrison:2011mb}.

\item The consistency conditions derived do not necessarily hold in general for supergravity theories coupled to non-trivial SCFTs. Examples of 6D supergravity theories with a strongly coupled sector which violate the aforementioned consistency conditions have been constructed in F-theory \cite{DelZotto:2014fia}. The constraints will hold at general points on the tensor branch of these theories where tensor multiplets take general vev's. At the superconformal points themselves the very definition of $\Lambda_S$ becomes problematic.

\end{itemize}

\section{Derivation of the new constraints}

\label{s: new}

\subsection{Local anomaly cancellation conditions}\label{ss: local}

In this section, we derive new constraints on the inner-products of anomaly polynomials from the local anomaly cancellation conditions.

\paragraph{Non-abelian anomaly coefficients} We start with Constraint \ref{Constr_a_char}, which reads
\be\label{2Z}
a \cdot b_i + b_i \cdot b_i \in 2\bZ
\ee
and holds for every simple summand $\mathfrak{g}_i$. \eqref{2Z} can be derived using the conditions
\bea\label{nonab}
a \cdot b_i  &=
{1 \ov 6} \left(A_{\text{Adj}_i}
- \sum_{\fR_i} x_{\fR_i} A_{\fR_i}\right) \,, \\
0&= B_{\text{Adj}_i}
- \sum_{\fR_i} x_{\fR_i} B_{\fR_i} \,, \\
b_i \cdot b_i  &=
{1 \ov 3} \left(-C_{\text{Adj}_i}
+ \sum_{\fR_i} x_{\fR_i} C_{\fR_i}\right) \,.
\eea
We show in Appendix \ref{AppRelMod12} that for any representation $\fR_i$ of $\fg_i$,
\be\label{12Z}
\chi_{\fR_i} := 2 C_{\fR_i} - A_{\fR_i} + \eta B_{\fR_i}
\in 12 \bZ \,,
\ee
where $\eta$ is a certain constant independent of the representation $\fR_i$. By taking a weighted sum of the three equations of \eq{nonab}, we arrive at equation \eq{2Z}.%
\footnote{Note that for localized matter in F-theory, $\frac{1}{12}\chi_{\fR_i}$ has an interpretation as the arithmetic genus associated to the singularity that the matter representation $\fR_i$ is localized on \cite{KPT,Grassi:2000we, Grassi:2011hq,Morrison:2011mb, Klevers:2017aku}. We thank Wati Taylor for pointing this out to us.}

\paragraph{Abelian anomaly coefficients} Recall that the abelian anomaly coefficients satisfy
\bea\label{u1 anom}
a \cdot b_{IJ} &= -{1 \ov 6} \sum_{q_I,q_J} q_I q_J x_{q_I,q_J} \\
b_i \cdot b_{IJ} &= \sum_{\fR,q_I,q_J} A_{\fR_i} q_I q_J x_{\fR,q_I,q_J} \\
b_{IJ} \cdot b_{KL} + b_{IK} \cdot b_{JL} + b_{IL} \cdot b_{JK}
 &= \sum_{q_I,q_J,q_K,q_L} q_I q_J q_K q_L x_{q_I,q_J,q_K,q_L} \,.
\eea

Let $\Lambda_\fR$ be the set of weights of the representation $\fR_i$, $d^{\fR_i}_w$ be the multiplicity of the weight $w$ in $\fR_i$, and $H_\theta$ be the coroot associated to the highest root $\theta$ of $\mathfrak{g}_i$. By the arguments of Appendix \ref{AppRelMod12},
\be
A_{\fR_i} = {1 \ov 2} \sum_{w \in \Lambda_{\fR_i}} d^{\fR_i}_w (w(H_\theta))^2
= \sum_{w \in \Lambda_{\fR_i}^+} d^{\fR_i}_w n_w^2 \;, \quad n_w \in \mathbb{Z}
\ee
is an integer, where $\Lambda_{\fR_i}^+$ is the subset of weights such that $w(H_\theta) > 0$. Recall that the field strength $F_I$ was normalized such that $F_I = 2\pi c_1^I$, thus imposing that all charges $q_I$ are integers. Thus it follows from equations \eq{u1 anom} that
\be
6 a\cdot b_{IJ}\,, \quad {b_i} \cdot b_{IJ} \,, \quad b_{IJ} \cdot b_{KL} + b_{IK} \cdot b_{JL} + b_{IL} \cdot b_{JK}
\ee
are integers. This proves Constraints \ref{Constr_2}, \ref{Constr_3} and \ref{Constr_4}

We may get a little bit more out of the local anomaly constraints involving abelian anomaly coefficients using the fact that the charges $q_I$ are integral. As $n^4-n^2$ is divisible by 12 for any $n \in \bZ$, we have
\be
b_{II}^2 +2 a \cdot b_{II} = {1 \ov 3} \sum_{q_I} x_{q_I} (q_I^4 - q_I^2) \in 4 \bZ \,,
\ee
which is Constraint \ref{Constr_5}. As $n^3-n$ is divisible by 6 for any $n \in \bZ$, we have
\be
b_{II}\cdot b_{IJ} +2 a \cdot b_{IJ} = {1 \ov 3} \sum_{q_I,q_J} x_{q_I,q_J} (q_I^3 - q_I)q_J \in 2 \bZ
\ee
which is Constraint \ref{Constr_6}. Similarly, we can get
\be
b_{II}\cdot b_{JJ} + 2b_{IJ}\cdot b_{IJ} +6 a\cdot b_{IJ}
= \sum_{q_I,q_J} x_{q_I,q_J} (q_I^2 q_J^2 - q_I q_J) \in 2\bZ\,,
\ee
which is Constraint \ref{Constr_7}. Constraint \ref{Constr_8} is derived along the same lines. The derivation is however lengthy and appears in Appendix \ref{ap:E}.

\subsection{Completeness hypothesis and global anomaly cancellation conditions} \label{ss: global}

We now turn to constraints imposed by global anomalies. A global anomaly is the change in the path integral measure under large gauge transformations or diffeomorphisms which cannot be smoothly connected to the identity. As explained previously, global gauge anomalies have been used to show the integrality of inner-products involving the non-abelian anomaly coefficients of gauge algebras of rank $\leq 2$ \cite{Bershadsky:1997sb, SuzukiTachikawa, KMT2}.

In general one should worry about nontrivial components of the automorphism group of general principal $G$ bundles on general six-dimensional spacetimes. In this section we will only consider spacetimes of the form $S^2 \times S^4$. Since we need to consider a topologically non-trivial manifold with fluxes  we assume the completeness hypothesis.

\subsubsection{A remark on the influence of the global structure of the gauge group.}

Although global anomalies a priori depend on the global form of the gauge group, the constraints we will derive in the present section depend only on the corresponding Lie algebra, for the following reason.

We have a fibration $\Gamma \rightarrow \tilde{G} \rightarrow G$, where $\Gamma$ is discrete. The long exact sequence of homotopy groups associated to the fibration above ensures that $\pi_4(G) \simeq \pi_4(\tilde{G})$. A gauge transformation on $S^4$ associated to a generator of $\pi_4(\tilde{G})$ can be projected to $G$ to yield a gauge transformation associated to a generator of $\pi_4(G)$. Moreover, given a $G$-gauge field configuration that is also a $\tilde{G}$-gauge field configuration, the action of the gauge transformations above coincide, because it only involves the adjoint actions of $G$ or $\tilde{G}$ on the corresponding Lie algebra $\mathfrak{g}$, and the latter coincide.

Now suppose there are fermions in a representation $\mathfrak{R}$ of $G$. By the above, we can see such fermions as $\tilde{G}$-valued fermions and study the constraints imposed by global anomalies. The constraints derived in this way are independent of the global form of the gauge group. In our case, this means that we can focus on a single non-abelian factor $\tilde{G}_i$ of the semisimple part $\tilde{G}_{\rm ss}$.

The discussion above does not exclude that stronger constraints could be derived by taking into account the global structure of the gauge group and considering $G$-gauge field configurations that are not $\tilde{G}$-gauge field configurations.

\subsubsection{$b_i \cdot b_{II} \in 2 \bZ$ for $\mathfrak{g}_i = \mathfrak{su}(2)$}

In Appendix \ref{ap:E}, we show that $b_i \cdot b_{II} \in 2 \bZ$ when there exists an $\mathfrak{su}(3)$ subalgebra of $\fg_i$ of embedding index 1. This includes all simple Lie algebras, except for $\mathfrak{sp}(n)$ (with $\mathfrak{su}(2) \simeq \mathfrak{sp}(1)$). We can still show the evenness of $b_i \cdot b_{II}$ for $\mathfrak{g}_i = \mathfrak{sp}(n)$ by using global anomaly cancellation on $M = S^2 \times S^4$ with a unit $\mathfrak{u}(1)_I$ flux along the $S^2$.

We start with the $n = 1$ case, i.e. $\mathfrak{g}_i = \mathfrak{su}(2)$, $\tilde{G}_i = SU(2)$. Let us recall some points from Witten's original analysis of the global $SU(2)$ anomaly \cite{Witten:1982fp}.

Witten considers Weyl fermions on $S^4$ valued in the fundamental representation of an $SU(2)$ flavor symmetry, i.e. valued in the representation $\mathbf{2_L} \otimes \mathbf{2}$ of ${\rm Spin}(4) \times SU(2)$. A sufficient criterion for the absence of global anomalies is the possibility of writing a mass term: one can then use Pauli-Villars regularization to deduce the absence of anomalies. A mass term would correspond to a antisymmetric tensor in $\mathbf{2_L} \otimes \mathbf{2}$. (It has to be antisymmetric because of the Fermi statistics.) But there is no such tensor. However, if we have two Weyl fermions, they are valued in $\mathbf{2_L} \otimes (\mathbf{2} \oplus \mathbf{2})$, and there is an antisymmetric tensor in $\mathbf{2} \oplus \mathbf{2}$. This means that if there is an anomaly, it can at most be a global sign (i.e. order 2) anomaly.

The fourth-homotopy group $\pi_4(SU(2))$ of the $SU(2)$ group manifold is given by $\bZ_2$, and there is a global gauge transformation on $S^4$ represented by the generator $g$ of this $\bZ_2$. Witten showed the Weyl fermion does indeed suffer from a global sign anomaly under this global gauge transformation. More generally, a fermion in the representation $\fR_i$ picks up a phase of $\pi A_{\fR_i}$, where $A_{\fR_i}$ is defined in equation \eq{ABCE} \cite{Bar:2002sa}.

Let us denote the $SU(2)$ gauge field configuration on $S^4$ as $a_i$, and consider the effect of acting on $a_i$ by a global gauge transformation $U(g)$, corresponding to the generator $g$ of $\pi_4(SU(2))$, i.e.,
\be
a_i ~\rightarrow~a_i^{U(g)} \,.
\ee
Let us compute the phase that a six-dimensional Weyl fermion in the representation $\fR_i \otimes q_I$ of $\mathfrak{su}(2) \oplus \mathfrak{u}(1)_I$ acquires under $U(g)$. In order to compute this phase, we must compute a mod-two index of a seven-dimensional fermion in the following gauge configuration on $\cM^6 \times \mathbb{R} = S^2 \times S^4 \times \mathbb{R}$. There is a constant unit $\mathfrak{u}(1)_I$ flux along the $S^2$ component, and a smooth five-dimensional $\mathfrak{su}(2)$ gauge field along the $S^4 \times \mathbb{R}$ direction that interpolates between $a_i$ on $S^4$ at $-\infty$ of the $\mathbb{R}$-coordinate, and $a^{U(g)}_i$ on $S^4$ at $+\infty$. The phase is given by this index multiplied by $\pi$.

Rather than computing the mod 2 index directly, we can reduce to four dimensions and use Witten's results. Recall that the six-dimensional fermions in a hypermultiplet of representation $\fR$ are Weyl fermions in the representation $\fR\otimes \mathbf{4}_L \oplus \fR^* \otimes \mathbf{4}_L^*$ of $G \times {\rm Spin}(5,1)$ satisfying a symplectic Majorana constraint. In Lorentzian signature, complex conjugation of the chiral spinor representation does not change the chirality. The subscript $L$ is used to emphasize this point. When analytically continued to Euclidean signature, this spinor can effectively be treated as a single Weyl fermion in the representation $\fR \otimes \mathbf{4}_L$ of $G \times {\rm Spin}(6)$ (see, for example, \cite{Green:1984bx}). The reduction of the Weyl fermion in the representation $q_I \otimes \fR_i \otimes \mathbf{4}_L$ of $\mathfrak{u}(1)_I \oplus \fg_i \oplus \mathfrak{so}(6)$ on the two dimensional sphere with unit flux yields
\be
x \cdot (\fR \otimes \mathbf{2_L}) \oplus
y \cdot (\fR \otimes \mathbf{2_R}) \quad \text{where $x-y =q_I$}
\ee
on $S^4$. This is because $\mathbf{4}_L$ decomposes into
\be
\mathbf{4_L} \rightarrow (+1,\mathbf{2_L}) \oplus (-1,\mathbf{2_R})
\ee
under the ${\rm Spin}(2) \times {\rm Spin}(4)$ subgroup of ${\rm Spin}(6)$ and $\text{index} (i \slashed{D}) = q$ for the two-dimensional Dirac operator of a charge $q$ fermion on a sphere with unit flux. According to \cite{Witten:1982fp}, each of these Weyl fermions picks up a phase of $\pi A_{\fR_i}$ under $U(g)$, so the phase picked up by the six-dimensional $\fR_i \otimes q_I$ fermion under $U(g)$ is given by
\be\label{phase}
\varphi_{\fR_i \otimes q_I} = \pi q_I A_{\fR_i} \,.
\ee
The anomaly is therefore present only for the representations having $A_{\fR_i}$ odd. \eqref{phase} implies that
\be
\sum_{q_I,\fR_i} x_{\fR_i,q_I} q_I A_{\fR_I}
\ee
is even, so
\be
\label{EqConstrbibIIsu2}
\sum_{q_I,\fR_i} x_{\fR_i,q_I} q_I^2 A_{\fR_I} = b_i \cdot b_{II} \in 2\bZ
\ee
for $\mathfrak{g}_i = \mathfrak{su}(2)$.

Note that standard Lie theory (see for instance 13.2.4 of \cite{CFT1997}) allows us to compute $A_{\fR_i}$ when $\mathfrak{g}_i = \mathfrak{su}(2)$ as
\be
A_{\fR^{(n)}} = \frac{{\rm dim} (\fR^{(n)}) (n\omega, n\omega + 2\rho)}{{\rm dim} (\mathfrak{su}(2))} = \frac{n(n+1)(n+2)}{6}\;,
\ee
where $\omega$ is the fundamental weight of $\mathfrak{su}(2)$, $\fR^{(n)}$ the representation with highest weight $n\omega$ (and spin $n/2$) and $\rho$ is the Weyl vector. We see that if $n$ is even, then $A_{\fR^{(n)}}$ is automatically an even integer, so the global anomalies do not affect $SO(3)$ factors. The constraint \eqref{EqConstrbibIIsu2} is nevertheless independent from the global form of the gauge group.

\subsubsection{$b_i \cdot b_{II} \in 2 \bZ$ for $\mathfrak{g}_i = \mathfrak{sp}(n), n > 1$}

Likewise, for $\fg_i = \mathfrak{sp}(n)$, $\pi_4(Sp(n)) = \bZ_2$ \cite{Bott}. The proof of evenness of $b_i \cdot b_{II}$ presented for $\fg_i = \mathfrak{su}(2) \simeq \mathfrak{sp}(1)$ would straightforwardly be replicated for the $\mathfrak{sp}$ algebras, given that the phase acquired by a four-dimensional Weyl fermion of representation $\fR_i$ under the global gauge transformation given by the generator of $\pi_4(Sp(n))$ is also $\pi A_{\fR_i}$. We conclude this section by proving that this is indeed the case.

There is an embedding of the group manifold $Sp(n)$ into that of $Sp(n+1)$ where $Sp(n+1)$ can be seen as a $Sp(n)$ fibration over $S^{4n+3}$ \cite{MimuraToda}:
\be
Sp(n) \rightarrow Sp(n+1) \rightarrow S^{4n+3} \,.
\ee
The homotopy groups $\pi_4(S^{4n+3})$ and $\pi_3(S^{4n+3})$ being trivial for $n \geq 1$, the inclusion $Sp(n) \rightarrow Sp(n+1)$ induces a homomorphism between $\pi_4(Sp(n))$ and $\pi_4(Sp(n+1)) \cong \bZ_2$. The chain of these homomorphisms maps the generator $g$ of $\pi_4 (Sp(1)) = \pi_4 (SU(2))$ into the generator of $\pi_4(Sp(n))$. At the same time, this embedding induces an embedding of $\mathfrak{su}(2)=\mathfrak{sp}(1)$ into $\mathfrak{sp}(n)$ such that the fundamental representation $\mathbf{2n}$ of $\mathfrak{sp}(n)$ decomposes into%
\footnote{This is because at each embedding $Sp(n) \rightarrow Sp(n+1)$ is such that the $\mathbf{2n+2}$ representation of $Sp(n+1)$ decomposes into
\be
\mathbf{2n+2} = \mathbf{2n} \oplus \mathbf{1}\oplus \mathbf{1}\,.
\ee}
\be
\mathbf{2n} = \mathbf{2} \oplus \mathbf{1}^{\oplus (2n-2)} \,.
\ee
Thus, given that a representation $\fR_i$ of $\mathfrak{sp}(n)$ decomposes into
\be
\fR_i = \bigoplus_R d^{\fR_i}_R \cdot R
\ee
of the $\mathfrak{su}(2)$ subalgebra, we find that the phase induced by the global gauge transformation $g \in \pi_4 (SU(2)) \cong \pi_4(Sp(n))$ on the Weyl fermion in representation $\fR_i$ is given by
\be\label{spn phase}
\sum_R d^{\fR_i}_R \varphi_R \,,
\ee
where $\varphi_R$ is the phase induces on the Weyl fermion in the representation $R$ of $\mathfrak{su}(2)$. This phase, however, has already been shown to be
\be\label{su2 phase}
\varphi_R = \pi A^{\mathfrak{su}(2)}_R \,.
\ee
Meanwhile, we can show that
\be\label{spn su2}
A^{\mathfrak{sp}(n)}_{\fR_i}= \sum_R  d^{\fR_i}_R A^{\mathfrak{su}(2)}_R \,,
\ee
by taking an element $F \in \mathfrak{su}(2) \subset \mathfrak{sp}(n)$ and using the fact that the embedding index of $\mathfrak{su}(2) \subset \mathfrak{sp}(n)$ is 1:
\be
\tr_{\mathbf{2n}} F^2 = \tr_{\mathbf{2}} F^2 \,, \quad
\tr_{\fR_i} F^2 = \sum_{R} d^{\fR_i}_R \tr_{R} F^2 \,.
\ee
Equation \eq{spn su2} can be obtained by taking the ratio of these two equations and remembering that ${\rm tr}$ is equal to ${\rm tr}_\mathbf{2}$ for $\mathfrak{su}(2)$ and to ${\rm tr}_{\mathbf{2n}}$ for $\mathfrak{sp}(n)$.

From equation \eq{spn phase}, \eq{su2 phase} and \eq{spn su2}, we find that the phase associated with a Weyl fermion in the representation $\fR_i$ of $\mathfrak{sp}(n)$ under the generator of $\pi_4(Sp(n))$ is given by $\pi A^{\mathfrak{sp}(n)}_{\fR_i}$, as desired. This shows that $b_i \cdot b_{II} \in 2\mathbb{Z}$ for $\mathfrak{g}_i = \mathfrak{sp}(n)$.

\subsection{Implications of the string charge quantization condition and of the completeness hypothesis}
\label{ss: charge quantization}

Let us finally derive the constraints imposed by the string charge quantization and the generalized completeness hypothesis on the anomaly coefficients. It will be instructive first to derive constraints ignoring the global structure of the gauge group, and then refine them using the global structure.

\subsubsection{Constraints independent of the global structure of the gauge group}

As any $\tilde{G}$-bundle is also a $G$-bundle, we use $\tilde{G}$-bundles to derive constraints independent of the global form of the gauge group.

We take $M = \mathbb{CP}^3$, and denote by $H$ the generator of the integral cohomology ring of $\mathbb{CP}^3$. The first Pontryagin class of $\mathbb{CP}^3$ is given by $[p_1] = 4 H \cup H$ \cite{CP3}. We will compute string charges along $\Sigma_4$, the cycle dual to $H \cup H$.

There is a $U(1)$-bundle $L$ with first Chern class given by $H$. Using $L$, we construct in Appendix \ref{AppIntc2} $\tilde{G}$-bundles $B_{II}$ and $B_{IJ}$, $I \neq J$, satisfying
\be
\label{EqConstrABunCP3_1}
c_2^i(B_{II}) = c_2^i(B_{IJ}) = 0 \quad \mbox{for all } i \;,
\ee
\be
\label{EqConstrABunCP3_2}
c_1^K(B_{II}) = \delta_{KI} H \quad \mbox{for all } K \:,
\ee
\be
\label{EqConstrABunCP3_3}
c_1^K(B_{IJ}) = \delta_{KI} H + \delta_{KJ} H\quad \mbox{for all } K \:,
\ee
as well as $\tilde{G}$-bundles $B_i$ for which
\be
\label{EqConstrNABunCP3_1}
c_2^i(B_i) = - H \cup H \;, \quad c_2^j(B_i) = 0 \quad \mbox{for all } j \neq i \;,
\ee
\be
\label{EqConstrNABunCP3_2}
c_1^I(B_i) = 0 \quad  \mbox{for all } I \;.
\ee
By the completeness hypothesis, we can consider our supergravity theory on $\mathbb{C}P^3$, and there are supergravity field configuration where the gauge fields are connections on the bundles $B_{II}$, $B_{IJ}$ or $B_i$. The string charge quantization condition \eqref{scq 2} applied to the bundles $B_{II}$, $B_{IJ}$ and $B_i$ then immediately yields Constraint \ref{Constr_10}. As explained above, the latter can be reformulated as Constraint \ref{Constr_10}*.

\subsubsection{Constraints dependent on the global structure of the gauge group.}

\label{SecConstrDepGlobStGG}

Let us now turn to the constraints imposed on the anomaly coefficients by the string charge quantization condition applied to all $G$-bundles. These constraints will of course depend on the global form of $G$.

We stay on $\mathbb{C}P^3$. Recall that a cocharacter is an element of the Cartan subalgebra of $\mathfrak{g}$ that is mapped to the identity by the exponential map of $G$. In Appendix \ref{AppIntc2}, we construct $G$-bundles $B_x$ depending on a cocharacter $x$ of $G$. The bundles $B_x$ satisfy
\be
\label{EqConstrNABunCP3_1b}
c_2^i(B_x) = - \frac{1}{2} {\rm tr}|_{\mathfrak{g}_i}(x^2) H \cup H \;, \quad c_1^I(B_x) = x^I H \;,
\ee
Here it is important to bear in mind that the cohomology classes associated to these Chern forms are not \emph{a priori} integral
if the $G$-bundle does not lift to a $\tilde{G}$ bundle. In particular,   $x^I$ denotes the  component of the cocharacter $x$ along $\mathfrak{u}(1)_I \subset \mathfrak{g}$ and might well be fractional. The $\tilde{G}$-bundles $B_{II}$, $B_{IJ}$ and $B_i$ are also $G$ bundles and coincide with $B_x$ for suitable coroots $x$. However, when the coroot lattice is a proper sublattice of the cocharacter lattice, which is generally the case, the constraints deduced from the bundles $B_x$ are stronger. Indeed, applying again the string charge quantization condition \eqref{scq 2}, we find the additional constraints
\be
\label{EqConstrbiGlob}
- \frac{1}{2} \sum_i b_i {\rm tr}|_{\mathfrak{g}_i}(x^2) \in \Lambda_S \;, \quad - b_{IJ} x^I x^J  \in \Lambda_S \;, \quad - \frac{1}{2} b_{II} x^I x^I  \in \Lambda_S \;.
\ee
for all cocharacters $x$ of $G$. The constraint \eqref{EqConstrbiGlob} can be interpreted as the requirement that the bilinear form $b$ is even and $\Lambda_S$-valued when restricted to the cocharacter lattice of $G$. We deduce that $\frac{1}{2}b(x,x) \in \Lambda_S$ for all cocharacters $x$. As explained in Appendix \ref{AppIntc2}, such bilinear forms are actually in bijection with $H^4(BG; \mathbb{Z})$, so we find that
\be
a \in \Lambda_S \;, \quad \frac{1}{2}b \in H^4(BG;\mathbb{Z}) \otimes \Lambda_S \;,
\ee
which is Constraint \ref{Constr_11}.

\subsubsection{Example: $U(2)$}

\label{SecExamU2}

Suppose that $G = U(2) = (SU(2) \times U(1))/\mathbb{Z}_2$.  Then the coroots form a lattice $\sqrt{2} \mathbb{Z} \oplus \mathbb{Z}$. The cocharacter lattice, i.e. the elements of the Cartan subalgebra of $U(2)$ exponentiating to the identity, is generated by the coroots and the additional cocharacter $x_{-1} = \left( \frac{1}{\sqrt{2}}, \frac{1}{2} \right)$, corresponding to the group element
\be
\left( \begin{array}{cc} -1 & 0 \\ 0 & -1 \end{array} \right) \times -1 \in SU(2) \times U(1)
\ee
generating $\Gamma = \mathbb{Z}_2$. We see that although a line bundle always has a first Chern form with integer periods, there is a $U(2)$-bundle $B_{x_{-1}}$ that has a first Chern form with half-integer periods. Similarly, the periods of the second Chern form of $B_{x_{-1}}$ are generally quarters of integers (as in the case of $SO(3)$-bundles).

Writing $b_0$ for the component of $b$ along $\mathfrak{su}(2)$ and $b_1$ for its component along $\mathfrak{u}(1)$, the cocharacter $x_{-1}$ yields the constraints
\be
b_0 \in 4\Lambda_S \;, \quad b_1 \in 8 \Lambda_S \;.
\ee

\section{F-theory} \label{s: Ftheory}

\subsection{Review of 6D F-theory vacua}
\label{ss: Ftheory background}

Before deriving constraints on the anomaly coefficients of F-theory vacua, let us first review some well-known facts about F-theory compactifications. Most of what we review in this section can be found in standard references including---but not restricted to---\cite{Morrison:1996na, Morrison:1996pp, KMT1, KMT2, TaylorTASI, SeibergTaylor}.

The data of a geometric 6D F-theory background is given by an elliptically fibered Calabi-Yau threefold $X \ra B$ with Section \cite{Morrison:1996na, Morrison:1996pp, TaylorTASI}. More precisely, one must specify the base $B$ of the elliptic fibration, a rational surface \cite{Morrison:1996pp}, as well as sections $f$ and $g$ of $-4K$ and $-6K$, which are the Weierstrass coefficients of the elliptic fibration:
\be\label{Weierstrass}
y^2 = x^3 + fx + g \,, \quad
f \in \cO(-4K) \,, ~
g \in \cO(-6K) \,.
\ee
Here, $K$ is the canonical bundle of $B$. By writing the Weierstrass model, we have implicitly chosen a zero section $Z$ of the fibration. The base $B$ of the elliptic fibration is to be identified with this zero section. There are various generalizations of the standard
$F$-theory setup but we will not consider them here.
\footnote{For example, an elliptic fibration---as opposed to a genus-one fibration---is not always necessary for defining an F-theory vacuum. 
But we must consider the elliptic fibration to access the component of the gauge group connected to the identity \cite{Braun:2014oya, Morrison:2014era}.
Also, it  has been shown that an elliptic fibration $X$ that does not have a smooth Calabi-Yau resolution can be used for an F-theory compactification, as long as it has a maximal crepant projective (MCP) blowup \cite{MPT}. The MCP blowup $\tilde{X}$ of such an $X$ contains $\bQ$-factorial terminal singularities, which do not compromise the physical integrity of the effective theory of the compactification. Throughout this work, we always assume that $X$ has a smooth Calabi-Yau resolution---it would be interesting to understand how to extend the results of this section to F-theory compactifications on elliptic fibrations with singular MCP blowups.}

In order for the low-energy effective theory of the compactification on $X$ to be described by the massless multiplets reviewed in section \ref{s: review and summary}, $X$ must be resolvable into a smooth Calabi-Yau manifold without blowing up the base \cite{Morrison:1996pp}. A singularity of $X$ whose resolution requires blow-ups along the base signals the existence of a 6D SCFT in the low-energy effective theory. We thus assume the absence of such singularities throughout this section. As noted before, the low-energy constraints derived in this work do not apply to supergravity theories with strongly coupled sectors, and are violated in many F-theory compactifications that have non-trivial SCFTs in their low-energy description \cite{DelZotto:2014fia}.%

\paragraph{String charge lattice} The effective theory of the F-theory compactification on the Calabi-Yau threefold $X$ is a 6D $\cN=(1,0)$ supergravity theory. The strings in the 6D theory come from wrapping D3-branes on 2-cycles of $B$. Their charges are therefore the corresponding elements of $H_2(B;\bZ)$ and we can identify $\Lambda_S = H_2(B;\bZ)_{\rm free}$. (We will ignore the possible torsion charges in what follows.) The pairing on $\Lambda_S$ then coincides with the intersection pairing on $H_2(B;\bZ)_{\rm free}$, and the number of tensor multiplets is given by $T := {\rm rank} (H_2(B;\bZ)) - 1$. $\Lambda_S$ is unimodular due to Poincar\'e duality.

\paragraph{Gravitational anomaly coefficient} The gravitational anomaly coefficient $a$ is to be identified with the homology class (modulo torsion) of the divisor of the canonical bundle $K$ of $B$ \cite{Sadov:1996zm, Grassi:2000we}. By definition, $a$ is an element of $\Lambda_S$. Furthermore, $a$ is a characteristic element of $\Lambda_S$, i.e. we have $(a,x) = (x,x)$ modulo 2 for any $x \in \Lambda_S$. This follows from the fact that given a smooth divisor $D \subset X$, $K_D$ is a characteristic element of the intersection pairing on $D$. Thus for any divisor $C$ of $D$,
\be
\label{EqAdjForm}
C \cdot (K_D + C) = C \cdot K_C = 2 (g_C - 1) \;,
\ee
where $g_C$ is the arithmetic genus of $C$. We have used the adjunction formula $K_D|_C + C = K_C$. We see that $a$ is a characteristic element of $\Lambda_S$ by setting $D=B$.

\paragraph{Non-abelian gauge algebra} The non-abelian gauge algebra is determined by the codimension-one singular locus of the elliptic fibration \eq{Weierstrass}, also known as the discriminant locus. For each irreducible curve component $i$ of the discriminant locus, above which the singular fiber is not of type $I_1$ or $II$, there is an associated simple non-abelian gauge algebra $\mathfrak{g}_i$ determined by the singularity type and monodromy of the singular fiber along the curve \cite{Grassi:2000we, Grassi:2011hq}.

\paragraph{Resolution} We write $\hat{X}$ for a resolution of the singular fibration $X$.%
\footnote{While there are multiple smooth Calabi-Yau resolutions of $X$, these resolutions are related to each other by a sequence of flops \cite{Aspinwall:1993nu, Intriligator:1997pq}. For our discussion, it does not matter which resolution is chosen, since the different resolutions can be interpreted as different phases of the Coulomb moduli space of the five-dimensional gauge theory obtained by compactifying F-theory on $X \times S^1$ \cite{Morrison:1996xf, Intriligator:1997pq, Witten:1996qb}.}
$\hat{X}$ contains extra divisors, the \emph{fibral divisors} $T_a$, obtained by fibering irreducible components of singular Kodaira fibers (which are rational curves) over the discriminant locus  \cite{2001math.....12259W, intersection, MP}. Each divisor $T_a$ corresponds to an element of the coroot basis of the Cartan subalgebra of $\mathfrak{g}_{\rm ss}$ \cite{Grassi:2000we}. We will not distinguish in notation between the divisors of $X$ and the corresponding divisors of $\hat{X}$.

\subsection{New constraints on anomaly coefficients in F-theory}
\label{ss: Ftheory constraints}

Before deriving constraints that the anomaly coefficients of 6D F-theory vacua satisfy, let us first identify the geometric counterpart of the Weyl-invariant bilinear form $b$ defined in Section \ref{ss: charge quantization}. We begin by writing ${\rm Pic}(Y)$ for the Picard group of an algebraic variety $Y$, i.e. the group of divisors up to linear equivalence. Let us consider the following pairing \cite{2001math.....12259W}:
\be
\begin{aligned}
& \tilde{b} : {\rm Pic}(\hat{X}) \times {\rm Pic}(\hat{X}) \rightarrow {\rm Pic}(B) \\
&\tilde{b}(D_1, D_2) :=  -\pi (D_1 \cdot D_2) \;,
\end{aligned}
\ee
where $\pi$ is the projection map from $\hat{X}$ to $B$.

Given a divisor $D$, let $[D]$ denotes its homology class modulo torsion. By taking the homology classes modulo torsion of the arguments and of the output, this pairing descends to a pairing $b$ on $H_4(X;\mathbb{Z})_{\rm free}$ valued in $H_2(B;\mathbb{Z})_{\rm free} \simeq \Lambda_S$:
\be\label{pairing b}
\begin{aligned}
& b : H_4(\hat{X};\mathbb{Z})_{\rm free} \times H_4(\hat{X};\mathbb{Z})_{\rm free} \rightarrow \Lambda_S \\
&b([D_1], [D_2]) := [\tilde{b}(D_1, D_2)] = -\pi_\ast([D_1] \cdot [D_2])\;,
\end{aligned}
\ee
where we used the relation between the intersection pairing on divisors and on homology classes, and $\pi_\ast$ denotes now the push-forward in homology. We have used the dot to denote both the intersection pairing between divisors and the intersection pairing between homology classes. The formula above for $b$ makes it clear that it is well-defined. When applying $b$ to the homology classes $[D_1]$, $[D_2]$ associated to divisors $D_1$ and $D_2$, we will freely write $b(D_1, D_2)$.%
\footnote{The pairing $b$, when composed with the Shioda map for threefolds \cite{Shioda, 2001math.....12259W, intersection,MP} becomes the height pairing of rational sections of the elliptically fibered Calabi-Yau manifold. $b$ has been referred to a ``height pairing" in the literature for this reason.}

Let us recall the useful fact that for every element ${h} \in H_4(\hat{X}; \bZ)$, there exists a divisor $D$ such that $[D] = {h}$. Due to the Lefschetz theorem on $(1,1)$-classes, we know that the first Chern class map ${\rm Pic}(\hat{X}) \xrightarrow{c_1} H^{1,1}(\hat{X}) \cap  H^{2}(\hat{X};\bZ)$ is surjective \cite{GH}. This map is obtained by taking the Poincare dual of the homology class of the divisor being mapped \cite{GH}. But we know that $H^{1,1}(\hat{X}) =  H^{2}(\hat{X};\bC)$ for a Calabi-Yau threefold $\hat{X}$, the classes $H^{2,0}(\hat{X})$ and $H^{0,2}(\hat{X})$ being empty. Thus we find that the map $D \mapsto [D]$ from ${\rm Pic}(\hat{X})$ to $H_4 (\hat{X};\bZ)$ is surjective.%
\footnote{In the F-theory literature, the term ``divisor" is thus often used interchangeably with the term ``homology class."}

Let $\{e_\alpha\}$ be a system of generators of $H_2(B;\bZ)_{\rm free}$, $\alpha = 0,...,T$, and write $\Om_{\alpha \beta}$ for the corresponding matrix elements of the intersection pairing on $H_2(B;\mathbb{Z})_{\rm free}$. Let us pick a divisor representative $E_{\alpha}$ for each $e_\alpha$. As in the Calabi-Yau threefold case, this is possible because $B$ is an algebraic surface with $h^{2,0} = 0$ \cite{KMT2}. Let us now consider the \emph{vertical divisors} of $\hat X$ defined by
\be
B_\alpha = \pi^{-1} (E_\alpha) \,.
\ee
We have
\be
[B_\alpha \cdot B_\beta] = \Om_{\alpha \beta} f \,,
\ee
where $f \in H_2(\hat{X};\bZ)_{\rm free}$ is the homology class of the generic fiber in $\hat{X}$. The pushforward map can then be written
\be
\pi_\ast([C]) = [\pi_\ast(C)] = (C \cdot B_\alpha) e^{\alpha} \,.
\ee
for $C$ any degree 2 algebraic cycle on $\hat X$. Here $e^\alpha = \Omega^{\alpha \beta} e_\beta$ and $\Omega^{\alpha \beta}$ are the matrix elements of $\Omega^{-1}$. As explained earlier, the intersection pairing on $H_2(B;\mathbb{Z})_{\rm free}$ is unimodular, and thus $\Omega^{-1}$ has integer entries and preserves the integrality properties of the classes. Indices $\alpha, \beta,...$ will always be lowered and raised with $\Omega$.

Following Shioda \cite{Shioda}, we consider the abelian group $\cH$, which is the orthogonal complement with respect to $b$ in $H_4(\hat{X};\bZ)_\free$ of the space spanned by the homology class of the zero section $[Z]$ and the classes $[B_\alpha]$:
\be
\cH = \vev{[Z], [B_0], ...\,,[B_{T}]}^\perp \cap H_4(\hat{X};\bZ)_\free \,.
\ee
For a divisor $D$, $[D]$ being an element of $\cH$ is equivalent to the conditions
\be
[D] \cdot f = 0 \,, \quad
[D|_Z] = 0 \,.
\ee
Here $D|_Z$ denotes the restriction of $D$ to the zero section $Z$, i.e. the cycle $D \cdot Z$, seen as a divisor of $Z$. The former follows from the fact that
\be
b(D,B_\alpha) = [D \cdot B_\alpha \cdot B_\beta] e^\beta
 = ([D] \cdot f) e_\alpha
\ee
and the latter follows from
\be
b(D,Z) =  [D \cdot Z \cdot B_\alpha] e^\alpha
= [D|_Z \cdot B_\alpha|_Z]  e^\alpha
= ([D|_Z] \cdot e_\alpha) e^\alpha
= [D|_Z] \,.
\ee
By construction, $[T_a] \in \cH$. We write $\cT := \vev{[T_a]}_\bZ$ for the subgroup of $\cH$ spanned by the homology classes of the fibral divisors.

As derived in \cite{intersection}, and will be further elaborated later on, the Cartan subalgebra of the gauge algebra $\fg$ can be identified with the vector space
\be
\cH \otimes \bR = \vev{[Z], [B_0], ...\,,[B_{T}]}^\perp \otimes \bR  \,.
\ee
The subspace $\cT \otimes \bR$ of this space is then identified as the Cartan subalgebra of $\fg_{\rm ss}$. We will show that $b|_{\cH \otimes \bR}$ is to be identified with the Weyl-invariant bilinear form $b$ defined in Section \ref{ss: charge quantization} by showing its relation to the non-abelian \eq{bi Ftheory} and abelian \eq{bIJ Ftheory} anomaly coefficients.

But before we do so, let us first prove an important property of the pairing $b$, when restricted to $\cH$.

\paragraph{Evenness of $b|_{\cH}$}
Let us now show that $b$ restricted to $\cH$ is even, that is,
\be
b(h,h) \in 2 \Lambda_S \quad \text{for all}~h \in \cH \,.
\ee
Let us take a divisor $D$ such that $[D] = h$. This statement is then equivalent to the claim that
\be\label{basic}
h \cdot h \cdot [B_\alpha]  = [D \cdot D \cdot B_\alpha]  \in 2 \bZ
\ee
for all $B_\alpha$. To prove it, we first note that
\be
\label{EqDBBZero}
[D \cdot B_\alpha \cdot B_\alpha] = \Omega_{\alpha\alpha} \: [D] \cdot f = 0
\ee
since $[D] = h \in \cH$. We then decompose $D$ onto a basis $D_i$ of effective divisors: $D = \sum_i n_i D_i$, $n_i \in \mathbb{Z}$. Bertini's theorem ensures that the $D_i$'s each admit a smooth linearly equivalent divisor, so we can assume that the $D_i$'s are all smooth. We then have
\be\label{individual}
\begin{aligned}
\, [D \cdot D \cdot B_\alpha] \, & = [D \cdot D \cdot B_\alpha] + [D \cdot B_\alpha \cdot B_\alpha] \\
& \equiv \sum_i n_i (D_i \cdot D_i \cdot B_\alpha + D_i \cdot B_\alpha \cdot B_\alpha) \quad ({\rm mod} ~ 2)\\
& = \sum_i n_i (D_i |_{D_i} \cdot B_\alpha|_{D_i} + B_\alpha|_{D_i} \cdot B_\alpha|_{D_i}) \\
&  = \sum_i n_i (K_{D_i} \cdot B_\alpha|_{D_i} + B_\alpha|_{D_i} \cdot B_\alpha|_{D_i}) \equiv 0  \quad
(\text{mod}~2) \,.
\end{aligned}
\ee
$|_{D_i}$ denotes the restriction of a divisor to $D_i$, yielding a divisor of $D_i$, and $K_{D_i}$ is the canonical divisor of $D_i$. %
In the first and second steps, we used \eqref{EqDBBZero} and dropped the even cross terms. We restricted to $D_i$ in the third step and used the definition of the canonical class in the fourth step. In the last step, we used the fact that $K_{D_i}$ is a characteristic element for the intersection pairing on $D_i$, as shown in \eqref{EqAdjForm}.

\paragraph{Non-abelian anomaly coefficients}
We can decompose the restriction of $b$ to $\cT$ along the irreducible components of the discriminant locus. The pairing $b$ then has the following form:
\be\label{bi Ftheory}
b|_{\cT} = \bigoplus_i \left( \cC_i \otimes b_i \right)
\ee
where $\cC_i$ is the normalized inner-product matrix of simple coroots of $\mathfrak{g}_i$ and $b_i \in \Lambda_S$. The explicit form of $\cC_i$ for each gauge algebra can be found in appendix A of \cite{intersection}. As the notation suggests, $b_i$ are exactly the non-abelian anomaly coefficients, so $b|_{\cT} = b_{\rm ss}$. This result is proven by studying the Chern-Simons terms of the dual M-theory description obtained by compactifying the F-theory background further down to five dimensions along a circle \cite{intersection,Ferrara:1996wv,Bonetti:2013cza}. Note that $\cC_i$ is always even, which is consistent with the fact that $b$ is even.

\paragraph{Abelian gauge algebra and abelian anomaly coefficients} Let us now consider $\cT^\perp$, which is defined to be the orthogonal complement of $\cT$ with respect to $b$ in $\cH$. The vector space $\cT^\perp \otimes \bR$ is to be identified with the abelian subalgebra $\mathfrak{g}_a$ \cite{intersection}. This has been shown by examining the Chern-Simons term of the five-dimensional theory obtained by compactifying the F-theory background along a circle and using the fact that none of the massless vector multiplets of the theory can be charged under the abelian gauge fields. The elements of $H_4(X;\bZ)_\free$ corresponding to $\mathfrak{u}(1)$ gauge fields (with integrally quantized charges) are then to be identified with the $r$ basis elements $s_I$ $(I=1,\cdots, r)$ of $\cT^\perp$. The abelian anomaly coefficients are given by ((3.43) of \cite{intersection}):
\be\label{bIJ Ftheory}
b_{IJ} = b(s_I ,s_J) \,.
\ee
The abelian anomaly coefficient matrix $b_{\rm a}$ is therefore identified with $b$ restricted to $\cT^\perp \subset \cH$. Since $b$ is even on $\cH$, $b_{\rm a}$ is even as well.

The relations \eq{bi Ftheory} and \eq{bIJ Ftheory} imply that $b$ is the Weyl-invariant pairing on the Cartan subalgebra of $\fg$ that defined the anomaly coefficients. The previous discussion also shows that the coroot lattice of the gauge algebra  
contains $ \cT \oplus \cT^\perp$ but in fact we have 
\be\label{coroot lattice}
\Lambda_\text{CR} = \cT \oplus \cT^\perp \, . 
\ee
To see this note that the coroot lattice of $G$ is the cocharacter lattice of $\widetilde{G}$ and the latter has an 
orthogonal decomposition in terms of the cocharacter lattice of the abelian and nonabelian parts. 
Since this lattice lies within $\cH$, it follows that for any element $D$ in the coroot lattice, $b(D,D) \in 2 \Lambda_S$. We thus arrive at the conclusion
\be
{1 \over 2} b \in H^4 (B\tilde{G} ; \bZ) \otimes \Lambda_S \,.
\ee

\paragraph{The global form of the gauge group}
Using the pairing $b$, we may further deduce the global form of the gauge group, much like in the manner that the intersection pairing of the K3 manifold has been used to understand the global form of the gauge group of eight-dimensional F-theory vacua \cite{Aspinwall:1998xj}. Recall that we have used $b$ to define the lattice
\be\label{cocharacter 2}
\cH = \vev{[Z], [B_0], ...\,,[B_{T}]}^\perp \cap H_4(\hat{X};\bZ)_\free \,.
\ee
We propose that this lattice $\cH$ should be identified with the cocharacter lattice $\Lambda^G_\text{CC}$ of the gauge group $G$. We note that $\cH$ has been implicitly studied and constructed using methods complementary to ours \cite{Mayrhofer:2014opa, Cvetic:2017epq}.%
\footnote{In these works, the Shioda map for elliptically fibered threefolds is used to identify certain linear combinations of the basis of the coroot lattice \eq{coroot lattice} with fractional coefficients that lie within $H_4(\hat{X};\bZ)_\free$. These integral homology classes are then used to build up the cocharacter lattice of $G$.}

We can prove that $\cH = \Lambda^G_\text{CC}$ by compactifying F-theory on $M' \times S^1 \times X$ and viewing the background as one obtained by compactifying M-theory on $M' \times \hat{X}$. The vector fields coming from reducing the M-theory three-form along elements of $H^2(\hat{X}; \bZ)_\free$ Poincar\'e dual to $H_4(\hat{X}; \bZ)_\free$ are gauge fields whose field strengths have periods valued in $H_4(\hat{X}; \bZ)_\free$.%
\footnote{Recall that the Poincar\'e dual element $\check{c} \in H_4(\hat{X}; \bZ)_\free$ of $c \in H^2(\hat{X}; \bZ)_\free$ is defined by its pairing with elements of $H_2$, i.e., the identity of the linear maps
\be
\check{c} \,\cdot = \vev{c,} \,,
\ee
where $\cdot$ is the intersection pairing and $\vev{,}$ is the canonical pairing between homology and cohomology elements.} \footnote{  
The reader may worry about a possible half-integral flux quantization of the M-theory C-field \cite{Witten:1996md}. Such fractional flux quantization does not arise in the present setup for the following reason. First we are looking exclusively at M-theory spacetimes of the form $Y = M' \times \hat{X}$. The shift in the flux quantization is governed by the 4th Stiefel-Whitney class $w_4(TY)$. As both $M'$ and $\hat{X}$ are spin, their second SW classes vanish and we can write $w_4(TY) = w_4(TM') + w_4(T\hat{X})$. Moreover, we have $w_4 = w_4 + w_2^2$ for both $TM'$ and $T\hat{X}$, which means that $w_4$ coincides with the degree 4 Wu classes of these two bundles. Now we can use the fact that the degree $p$ Wu class vanishes on any manifold of dimension strictly lower than $2p$ to conclude that $w_4(TY) = 0$ and that the C-field fluxes are integrally quantized.} As usual, we refer to such a gauge field as one obtained by compactifying the three-form along a four-cycle, or equivalently, as one that ``corresponds to" the four-cycle, with the understanding that the three-form is being compactified along the Poincar\'e dual cohomology element of that four-cycle.

This M-theory background, at a generic point in the five-dimensional moduli space, can be understood as being in the Coulomb phase of a five-dimensional gauge theory. The five-dimensional gauge fields obtained from the reduction of the C-field along $Z$ or $B_\alpha$ are respectively the Kaluza-Klein gauge fields associated to the metric and to the tensor gauge fields, so only the five-dimensional gauge fields obtained from elements of $\cH$ correspond to six-dimensional gauge fields \cite{intersection,Ferrara:1996wv,Bonetti:2013cza}. We then find that the magnetic fluxes along any given 2-cycle in the six-dimensional theory are precisely labelled by the lattice $\cH$. We obtain the five-dimensional magnetically charged strings corresponding to the six-dimensional self-dual strings by wrapping 5-branes around $4$-cycles in $\cH$.

We may now obtain the discrete group $\Gamma$ associated to the global form of F-theory gauge group $G$ (see \eqref{EqGlobFormGaugeGroup}). Since we have shown that the lattice
\be
\cT \oplus \cT^\perp
\ee
is the coroot lattice, and that $\cH$ is the cocharacter lattice of $G$, $\Gamma$ is
\be
\Gamma = \Lambda^G_\text{CC} / \Lambda_\text{CR}
= \cH / [ \cT \oplus \cT^\perp ]  \,.
\ee
The global form of (the connected part of) the $F$-theory gauge group is determined by the action of $\Gamma$ on the cover $\tilde{G}$. But the latter is the exponential of the additive action of $\cH$ on the Cartan subalgebra of $G$, which can be deduced from the embedding of $\cT \oplus \cT^\perp$ into $\cH$.

\paragraph{Consistency conditions on the pairing $b$} 

We have shown previously that for all elements $h$ of $\cH$, $b(h, h) \in 2 \Lambda_S$, and now have shown that $\cH$ is to be identified with $\Lambda^G_\text{CC}$. It follows that ${1 \over 2} b \in H^4(BG; \bZ) \otimes \Lambda_S$ in F-theory. We thus find that the anomaly coefficients of F-theory compactifications satisfy all consistency conditions derived in the previous sections.

\paragraph{A consistency check}

As a consistency check on our picture of the global form of the F-theory gauge group, we verify that the electric charge lattice is the character lattice $\Lambda^G_\text{C}$ of $G$. 
The electric charge lattice of the $F$-theory compactification 
can be viewed as the sublattice of $H^4(\hat{X}; \bZ)_\free$ orthogonal to $[Z]$ and $[B_\alpha]$:
\be\label{gauge electric charge}
\Lambda^G_\text{C} =  \text{Ann} (\vev{[Z], [B_0], ...\,,[B_{T}]}) \cap H^4(\hat{X};\bZ)_\free \,,
\ee
where the {\it annihilator} of a subspace $W$ of $H_4(\hat{X}, \bR)$ is
\be
\text{Ann} (W) = \{ v ~:~ v \in H^4(\hat{X};\bZ)\,, ~ \vev{v,w} =0 ~\text{for all } w \in W \} \,.
\ee
In order for our results to be consistent, $\Lambda^G_\text{C}$ and $\Lambda^G_\text{CC}$, which we have defined independently, must be dual to each other. In pedagogical terms, given $\Lambda^G_\text{CC} = \cH$ and $\Lambda^G_\text{C}$ defined by equations \eq{cocharacter 2} and \eq{gauge electric charge} respectively, there must be bases $\{ h_x \}$ of $\cH$ and $\{ \tilde{h}^x \}$ of $\Lambda^G_\text{C}$ such that
\be
\vev{\tilde{h}^x, h_y} = \delta^x_y \,.
\ee

To prove that this is the case, we first show that $\cH \oplus \vev{[Z], [B_0], ...\,,[B_{T}]}_\bZ = H_4(\hat{X};\bZ)_\text{free}$, i.e., that the integral basis of $\cH$, along with $[Z]$ and $[B_\alpha]$, spans the entirety of the lattice $H_4(\hat{X};\bZ)_\text{free}$. This is equivalent to the statement that for any $h \in H_4(\hat{X};\bZ)_\text{free}$, there exist integers $z$, $n^\alpha$ such that
\be
h- z[Z] - n^\alpha [B_\alpha] \in \cH \,.
\ee
We can find the required $z$ and $n^\alpha$ by the intersection numbers
\be
z = h \cdot f\,, \quad
n^\alpha = (h - z[Z]) \cdot [Z] \cdot [B^\alpha] \,.
\ee
Recall that $[B^\alpha] = \Om^{\alpha \beta} [B_\beta]$ where $\Om^{\alpha \beta}$ is used to denote the components of the inverse of the intersection pairing $\Om$ on the unimodular lattice $H_2(B;\bZ)_\free$. Note that $[B^\alpha]$ are integral four-cycles, the lattice $H_2(B;\bZ)_\free$ being unimodular. $z$ and $n^\alpha$, being intersection numbers between integral classes are integers. Furthermore,
\be
(h - z[Z] - n^\alpha [B_\alpha]) \cdot f = 0  \,, \quad
(h - z[Z] - n^\alpha [B_\alpha]) \cdot [Z] \cdot [B_\beta] = 0
\ee
for all $\beta$. These equalities may be derived by using the identities
\be
[Z] \cdot f = 1 \,, \quad [B_\alpha] \cdot f = 0 \,, \quad
[Z] \cdot [B_\alpha]  \cdot  [B_\beta] = [Z \cdot B_\alpha  \cdot B_\beta] = \Om_{\alpha\beta}  \,.
\ee
Thus $(h - z[Z] - n^\alpha [B_\alpha])$ must be a member of $\cH$.

Let us now denote the basis of $\cH$ as $h_1, \cdots, h_R$, where $R$ is the total rank of the gauge algebra. Since
\be
\{ [Z], ~[B_0], \cdots,~ [B_T], ~h_1, \cdots,~h_R \}
\ee
form an integral basis of $H_4(\hat{X};\bZ)_\text{free}$, their dual cocycles
\be
\{ f, ~[E^0], \cdots,~ [E^T], ~\tilde{h}^1, \cdots,~ \tilde{h}^R \}
\ee
with respect to this basis set lies within---and furthermore, is an integral basis of---$H^4(\hat{X};\bZ)_\free$ by the duality between $H_4(\hat{X};\bZ)_\free$ and $H^4(\hat{X};\bZ)_\free$. This implies that
\be
\cH^* = \vev{ \tilde{h}^1 ,\cdots, \tilde{h}^R }
\ee
is precisely the sublattice of $H^4(\hat{X};\bZ)_\free$ that annihilate $[Z]$ and $[B_\alpha]$ with respect to the canonical pairing and thus
\be
\cH^* = \Lambda^G_{C} \,.
\ee
Meanwhile, by construction, $\cH^*$ is dual to $\cH$ with respect to the canonical pairing, i.e., the condition
\be
\vev{\tilde{h}^x , h_y} = \delta^x_y
\ee
is satisfied, as desired.

\paragraph{Completeness of particle spectrum in F-theory}

We can also check that the completeness hypothesis is satisfied in F-theory, in that the charges of the multi-particle spectrum of the six-dimensional F-theory backgrounds studied saturates the weight lattice $\Lambda^G_\text{C}$. We do so by first showing that this is the case in the five-dimensional M-theory background. The charged states in the five-dimensional background can be obtained by wrapping M2 branes along the elements of $H_2(\hat{X}; \bZ)_\text{free}$.  The states obtained by wrapping M2 branes on $c \in H_2 (\hat{X}; \bZ)_\free$, while not guaranteed to be single particle or BPS states, are nevertheless states with charge $(c \cdot D)$ under a gauge field obtained by compactifying the M-theory three-form along a four-cycle $D \in \cH \subset H_4(\hat{X};\bZ)_\free$. Since an M2 brane can wrap any element of $H_2(\hat{X}; \bZ)_\text{free}$, it can represent any element of the dual $\cH^\ast$ of $\cH$.%
\footnote{To avoid introducing further notation, we still use $\cH^*$ to denote the subgroup of $H_2(\hat{X}; \bZ)_\text{free}$ corresponding $\cH^* \subset H^4(\hat{X}; \bZ)_\text{free}$ under the isomorphism determined by Poincar\'e duality.} 
The five-dimensional charged states of the theory saturate the character lattice $\Lambda^G_{\rm C} = \cH^*$.

The five-dimensional states coming from the M2 branes wrapping elements of $\cH^* \subset H_2(\hat{X}; \bZ)_\text{free}$ are not charged under gauge fields corresponding to $[Z]$ or $[B_\alpha]$. Recall that from the point of view of the six-dimensional F-theory background, the gauge field corresponding to $[Z]$ is obtained by reducing the six-dimensional graviton, and those corresponding to $[B_\alpha]$ are obtained by reducing six-dimensional tensor fields \cite{intersection,Ferrara:1996wv,Bonetti:2013cza}. We thus see that these five-dimensional states cannot come from wrapping a string along the compactification circle, and must originate from a six-dimensional state living on a spatial slice of space-time. Furthermore, these states must have zero Kaluza-Klein momentum. As a consequence, we find that the charges of the six-dimensional spectrum under the Cartan subalgebra of the gauge group also saturate the character lattice $\Lambda^G_{\rm C}$.

\subsection{Summary of constraints}

We have shown that in F-theory, ${1 \over 2} b \in H^4(BG; \bZ) \otimes \Lambda_S \subset H^4(B\tilde{G}; \bZ) \otimes \Lambda_S $. We have also noted that the gravitational anomaly coefficient $a$ is a characteristic vector in F-theory, i.e.
\be
a \cdot x + x \cdot x \in 2\Lambda_S \mbox{ for all } x \in \Lambda_S \,,
\ee
it being the canonical class of the base manifold. We do not know whether there are low-energy constraints in six-dimensional supergravity imposing this constraint on the gravitational anomaly coefficient.

\section{Summary and future directions} \label{s: future}

Using local and global anomaly cancellation conditions, we were able to derive new constraints on the anomaly coefficients of 6D $\cN =(1,0)$ supergravity theories, appearing as Constraints \ref{Constr_a_char}-\ref{Constr_8} in Section \ref{SecNewConstrRes}. Among these constraints, the most interesting one is maybe
\be
a \cdot b_i + b_i \cdot b_i \in 2 \bZ
\ee
which says that $a$ is a characteristic vector of the lattice generated by the non-abelian anomaly coefficients $\{b_i\}$. Indeed, this is the only constraint that is not implied by the generalized completeness hypothesis. This result should be compared to the fact that in F-theory, the gravitational anomaly coefficient $a$ has to be a characteristic vector of $\Lambda_S$, i.e.
\be
a \cdot x + x \cdot x \in 2\bZ
\ee
for any $x \in \Lambda_S$.

We showed that the generalized completeness hypothesis, when combined with the string charge quantization condition, yields a set of constraints on the anomaly coefficients that can be elegantly summarized as
\be
a \in \Lambda_S \;, \quad \frac{1}{2}b \in H^4(BG;\mathbb{Z}) \otimes \Lambda_S \;.
\ee
Unlike Constraints \ref{Constr_a_char}-\ref{Constr_8}, these constraints also hold in the presence of half-hypermultiplets and depend on the global structure of the gauge group. They are the weakest when the gauge group is a direct product of a simply connected semi-simple group with an abelian group. 

We have also shown that these constraints are all satisfied in F-theory backgrounds. In the process, we have been able to derive the cocharacter lattice $\Lambda^G_{\rm CC}$ of $G$:
\be
\Lambda^G_{\rm CC} = \cH = \vev{[Z], [B_0], ...\,,[B_{T}]}^\perp \cap H_4(\hat{X};\bZ)_\free \,,
\ee
where $\hat{X}$ is the smooth Calabi-Yau resolution of the elliptically fibered compactification manifold $X$, $Z$ is the zero section and $B_\alpha$ are the vertical divisors obtained by taking the preimage of representatives of the integral basis of the base homology group with respect to the projection map. The symbol $\perp$ denotes orthogonality with respect to the pairing $b$ defined in \eq{pairing b}. Defining $\cT$ to be the sublattice of $\cH \subset H_4(\hat{X};\bZ)_\free$ generated by the homology classes of the fibral divisors, and $\cT^\perp$ to be its orthogonal complement within $\cH$, we find that the coroot lattice of the gauge algebra is given by
\be
\Lambda_{\rm CR} = \cT \oplus \cT^\perp \,.
\ee
The global form of the gauge group is then determined by the embedding $\Lambda_{\rm CR} \subset \Lambda^G_{\rm CC}$. In particular, we can express the gauge group of the F-theory background as (see \eq{EqGlobFormGaugeGroup})
\be
G = (\tilde{G}_{\rm ss} \times G_a) / \Gamma \,,
\ee
where $\tilde{G}_{\rm ss}$ is a semi-simple simply connected group, $G_{\rm a} \simeq U(1)^r$ is an abelian group, and $\Gamma$ is a discrete abelian group. Then
\be
\Gamma = \cH / (\cT \oplus \cT^\perp) \;,
\ee
and its action on $\tilde{G} = \tilde{G}_{\rm ss} \times G_a$ is determined by the exponential of the action of $\cH$ on the Cartan subalgebra of $\tilde{G}$.

It is quite easy to construct supergravity theories that satisfy all the known low-energy constraints listed in this paper, but whose gravitational anomaly coefficient is not a characteristic vector. For example, a theory with $T=1$ with the quadratic form
\be
\Om =
\begin{pmatrix}
0&1\\1&0
\end{pmatrix}
\ee
on $\Lambda_S = \bZ^2$ with $a =(4,1)$ that has no gauge symmetry and $244$ neutral hypermultiplets satisfies all known low-energy consistency conditions. Nevertheless, $a$ is not a characteristic vector:
\be
a \cdot x + x \cdot x = 1 \notin 2 \bZ \;, \quad \mbox{for } x=(1,0)\;,
\ee
and this theory cannot be realized in F-theory. It would be interesting to understand if these theories violate any low-energy consistency conditions we have failed to discuss in this work.

We will present in a future publication a careful construction of the Green-Schwarz term \eqref{GS}. This construction  requires $a$ to be a characteristic element of $\Lambda_S$. A proof of uniqueness for this construction would then explain why $a$ has to be a characteristic vector, and would prove that the supergravity theory described above is inconsistent.

It is very striking that $H^4(BG;\bZ)$ appears naturally in formulating the quantization conditions for the 6d anomaly coefficients. 
This cohomology group is also the level of a 3d Chern-Simons gauge theory with gauge group $G$. Surely, this is not a coincidence. 
The natural way to try and understand the relation between these facts is to study the rational conformal field theories related to compactification of the 6d F-theory down to two dimensions. Carrying out the details of this suggestion seems well worthwhile. 

Another direction for future work would be to apply methods presented in this work to identifying the global form of the gauge group in four-dimensional F-theory compactifications.

\acknowledgments

We would like to thank Dave Morrison for discussions and Wati Taylor for useful comments on a draft. G.M. is supported by the DOE under grant DOE-SC0010008 to Rutgers University. S.M. is supported in part by SNSF grants No. 152812, 165666, and by NCCR SwissMAP, funded by the Swiss National Science Foundation. The work of D.S.P. has also been supported by DOE grant DOE-SC0010008.

\appendix

\section{Integrality properties of the second Chern class}

\label{AppIntc2}

In this appendix, we determine the integrality properties of the classes
\be
\label{EqDefc2}
c_2^i := -\frac{1}{8\pi^2} {\rm tr} F^2_i \;,
\ee
where $F_i$ is the component of the curvature of the gauge field along $\mathfrak{g}_i$, and tr is the trace in the adjoint representation of $\mathfrak{g}_i$ divided by twice the dual Coxeter number. This is the normalization of the Killing form ensuring that the long roots have length square 2.

The characteristic forms $c_2^i$, $i = 1,...,s$, generate a rank $s$ free subgroup $\Lambda_c \subset H^4(BG;\mathbb{R})$. In order to understand the integrality properties of $c_2^i$, we need to compare $\Lambda_c$ to $H^4_{\rm int}(BG;\mathbb{R})$, the subgroup of degree 4 characteristic forms of $G$-bundles with integral periods.

Taking inspiration from Section 4 in \cite{Dijkgraaf:1989pz}, we will proceed as follows. We will first characterize the elements of $H^4_{\rm int}(BG;\mathbb{R})$, by giving constraints on their values on a set of elementary $G$-bundles. We will then compute the value of $c^i_2$ on the same bundles, which will provide the desired characterization of $\Lambda_c$ with respect to $H^4_{\rm int}(BG;\mathbb{R})$.

\paragraph{Cartan tori and induced bundles} Let $\Gamma$ be the subgroup of the center of $\tilde{G}$ such that $G \simeq \tilde{G}/\Gamma$, as in \eqref{EqGlobFormGaugeGroup}. Let $T$ and $\tilde{T}$ be respectively maximal tori in $G$ and $\tilde{G}$. Recall that $\tilde{T}$ can be pictured as the quotient of the Cartan subalgebra of $\mathfrak{g}$ by the coroot lattice $\Lambda_{\rm CR}$. $T$ is a quotient of $\tilde{T}$, and therefore can be seen as the quotient of the Cartan subalgebra by a lattice $\Lambda^G_{\rm CC}$ containing the coroot lattice. $\Lambda^G_{\rm CC}$ is none other than the cocharacter lattice of $G$, defined as the group of homomorphisms from $U(1)$ to the Cartan torus $T$ of $G$. The elements of $\tilde{T}$ sent to $1 \in T$ form a finite discrete group canonically isomorphic to $\Gamma$, so $\Gamma \simeq \Lambda^G_{\rm CC}/\Lambda_{\rm CR}$. Note that the cocharacter lattice of $\tilde{G}$ can be identified as the coroot lattice $\Lambda_{\rm CR}$. As we will see, to study the integrality properties of $c_2^i$, it is sufficient to consider $G$-bundles induced from $T$-bundles.

Given a $U(1)$-bundle $L$, we can induce a corresponding $T$-bundle $K^T_{x,L}$, and therefore a $G$-bundle $K_{x,L}$ for each $x \in \Lambda^G_{\rm CC}$.

\paragraph{Degree 4 characteristic classes as invariant bilinear forms} The cohomology of $BT$ is generated by degree 2 elements, the first Chern classes, which are in bijection with a basis of the weight space of $\mathfrak{g}$.
$H^4(BT;\mathbb{R})$ is the symmetric tensor product of two copies of the weight space of $\mathfrak{g}$, and $H^4(BG;\mathbb{R})$ can be identified with the invariant subspace of $H^4(BT;\mathbb{R})$ with respect to the action of the Weyl group of $\mathfrak{g}$ (which coincides with the Weyl group of $\mathfrak{g}_{\rm ss}$). We see therefore that $H^4(BG;\mathbb{R})$ can be pictured as the spaces of Weyl invariant bilinear forms on the Cartan subalgebra of $\mathfrak{g}$.

\paragraph{Characterization of $H^4_{\rm int}(BG;\mathbb{R})$} We now prove that $H^4_{\rm int}(BG;\mathbb{R})$ is isomorphic to the subgroup of invariant bilinear forms $c$ on the Cartan subalgebra of $\mathfrak{g}$ such that $c(x,x) \in \mathbb{Z}$ for all $x \in \Lambda^G_{\rm CC}$.
Such a bilinear form $c$ can be expressed as an element of the symmetric tensor product of $(\Lambda^G_{\rm CC})^\ast$ with itself, where $(\Lambda^G_{\rm CC})^\ast = \Lambda^G_{\rm C}$ is nothing but the lattice of characters of $G$. Naturally, the characters of $G$ are homomorphisms from the Cartan torus of $G$ into $U(1)$. Meanwhile, $\Lambda^G_{\rm C} \simeq H^2(BT;\mathbb{Z})$ and $H^4(BT;\mathbb{Z})$ is the symmetric tensor product of $H^2(BT;\mathbb{Z})$ with itself. We now use the fact that the pull-back map $H^4(BG;\mathbb{Z}) \rightarrow H^4(BT;\mathbb{Z})$ associated to the inclusion $T \subset G$ is injective, with its image given by the Weyl invariant elements of $H^4(BT;\mathbb{Z})$. As $c$ is Weyl invariant by hypothesis, it corresponds to a unique element in $H^4(BG;\mathbb{Z})$. To conclude, we use the fact that $H^4(BG;\mathbb{Z})$ is free \cite{2016arXiv160202968H}, so it is isomorphic to $H^4_{\rm int}(BG;\mathbb{R})$. The chain of arguments above is valid in both directions.

\paragraph{Chern classes of $K_{x,L}$} Suppose $L$ has curvature $F_L$. Then the first Chern class of $K_{x,L}$ associated to the character $w$ is \be
\label{EqDefFirstChernClassWeight}
c_1^{(w)}(K_{x,L}) := \frac{1}{2\pi} w(x) F_L \;,
\ee
where $w(x)$ is the evaluation of the character $w$ on the cocharacter $x$. Note that for this class to be non-zero in cohomology, the character $w$ has to be Weyl invariant, i.e. it should lie in the abelian summand of the Lie algebra $\mathfrak{g}$. 

We can naturally identify $c_2^i$ with the bilinear form $-\frac{1}{2} {\rm tr}|_{\mathfrak{g}_i}$, which is fed $F_i/2\pi$ to yield \eqref{EqDefc2}. Therefore
\be
\label{EqC2Kx}
c_2^i(K_{x,L}) = -\frac{1}{2} {\rm tr}|_{\mathfrak{g}_i}(x^2) \left( \frac{F_L}{2\pi} \right)^2 \;.
\ee

\paragraph{Summary} We have shown that the degree 4 characteristic forms of $G$ bundles with integral periods are those corresponding to Weyl invariant bilinear forms $c$ on the Cartan subalgebra of $\mathfrak{g}$ such that $c(x,x) \in \mathbb{Z}$ for all $x \in \Lambda^G_{\rm CC}$. An equivalent condition is that $2c$ should be an even integral bilinear form. We have also shown that $c^i_2$ corresponds to the bilinear form $-\frac{1}{2} {\rm tr}|_{\rm \mathfrak{g}_i}$. This completely characterizes the integrality properties of $c^i_2$. In general, it has fractional periods, given by multiples of $-\frac{1}{2} {\rm tr}|_{\rm \mathfrak{g}_i}(x^2)$, for $x \in \Lambda^G_{\rm CC}$.

Note that if $\Gamma$ happens to be trivial, then $\Lambda^G_{\rm CC}$ coincides with $\Lambda_{\rm CR}$. But $-\frac{1}{2} {\rm tr}|_{\rm \mathfrak{g}_i}(x^2) = -1$ when $x$ is a coroot dual to a long root, and $-\frac{1}{2} {\rm tr}|_{\rm \mathfrak{g}_i}(x^2) = -2$ or $-3$ when $x$ is dual to a short root. Therefore $c^i_2$ has integral periods. This shows in particular that $c^i_2$ has integral periods if $G$ is simply connected (in which case there is no abelian component).

\paragraph{Gauge bundles on $\mathbb{C}P^3$} On $\mathbb{C}P^3$, $H^2(\mathbb{C}P^3;\mathbb{Z})$ is generated by a class $H$, and $H^4(\mathbb{C}P^3;\mathbb{Z})$ is generated by $H \cup H$. Let $L$ be a $U(1)$-bundle with first Chern class equal to $H$ and let $x$ be a coroot of $\mathfrak{g}_i$ dual to a long root. As $x$ is a coroot, $K_{x,L}$ is a $\tilde{G}$-bundle. Let $B_i := K_{x,L}$. By the discussion above, $B_i$ satisfies \eqref{EqConstrNABunCP3_1} and \eqref{EqConstrNABunCP3_2}. Now let $x_I$ be the $I$th canonical basis vector of $\mathfrak{g}_a \simeq \mathbb{R}^r$. $x_I$ is a coroot of $\mathfrak{g}$, and there is an associated $\tilde{G}$-bundle $B_{II} := K_{x_I,L}$. $B_{II}$ satisfies \eqref{EqConstrABunCP3_1} and \eqref{EqConstrABunCP3_2}. Similarly, define the $\tilde{G}$-bundle $B_{IJ} := K_{x_I + x_J, L}$, which satisfies \eqref{EqConstrABunCP3_1} and \eqref{EqConstrABunCP3_3}. Like any $\tilde{G}$-bundles, $B_{II}$ and $B_{IJ}$ are also $G$-bundles. Finally, given any cocharacter $x \in \Lambda^G_{\rm CC}$, let $B_x := K_{x,L}$. By the discussion above, the $G$-bundles $B_x$ satisfy \eqref{EqConstrNABunCP3_1b}.

\paragraph{Quantization of $c_2$ for adjoint groups} We compute here the quantization of $c_2$ for the adjoint simple Lie groups. The highest weights of adjoint group representations are always roots, so the character lattice coincides with the root lattice. Therefore the adjoint group cocharacter lattice is given by the coweight lattice, which is dual to the root lattice of the Lie algebra. Let $C$ be the Cartan matrix, defined by $C_{ij} = \frac{2(\alpha_i, \alpha_j)}{(\alpha_i, \alpha_i)}$, where $\alpha_i$ are the simple roots of the Lie algebra. This definition implies that the Killing form in the simple root basis is given  by the matrix $S = DC$, where $D_{ij} = \frac{1}{2} \delta_{ij} (\alpha_i, \alpha_i)$. Then the pairing of the coweight lattice in the dual basis is given by $S^{-1}$. It is a bit inconvenient to work with coweights. We can use the fact that the coweight lattice of a Lie algebra is the weight lattice of its Langlands dual up to a scaling factor. This factor can be determined from the fact that the coroots dual to long roots become the short roots of the Langlands dual. It is 1 for the simply laced Lie algebras, 2 for $B_n$, $C_n$ and $F_4$, and $3$ for $G_2$. We can use for $S^{-1}$ the explicit matrix form of the Killing form in the fundamental weight basis as listed for instance in Appendix 13.A of \cite{CFT1997} under the name ``quadratic form matrix", provided we exchange the $B$ and $C$ series and multiply it by the appropriate factor.

In order to find the minimal charge for $c_2$, we need to find the coweight with minimal norm. For a general lattice, this is a difficult problem, but we can again use the fact that we are working with the weight lattice of the Langlands dual. Recall that a weight is called dominant if it is a non-negative integer linear combination of the fundamental weights. Also, any weight is Weyl conjugate to a dominant weight, and the Weyl group preserves the weight norm. As $S^{-1}$ has positive entries only, this means that the weight of minimal norm has to be one of the fundamental weights. Switching back to coweights of the original Lie algebra, we list in Table \ref{t:QuantFracc2} all the fractional charges associated to these coweights.

\begin{table}[h]
\center
 \begin{tabular}{|c|ccccccccc|} \hline
 & $A_n$ & $B_n$ & $C_{n}$ & $D_n$ & $E_6$ & $E_7$ & $E_8$ & $F_4$ & $G_2$ \\ \hline
\multirow{3}{*}{$q$} & $\frac{k(n-k+1)}{2n+2}$ & $\frac{k}{2}$ & $\frac{n}{4}$ & $\frac{k}{2}$ & $\frac{2}{3}$ & $\frac{15}{4}$ & $1$ & $1$ & $1$ \\
& & & & $\frac{n}{8}$ & $\frac{5}{3}$ & $\frac{3}{4}$ & & &  \\
& & &  & $\frac{n}{8}$ & & $\frac{7}{4}$ & & &  \\ \hline
\multirow{3}{*}{$x$}  & $\alpha^\ast_k$ & $\alpha_k^\ast$ & $\alpha_n^\ast$ & $\alpha_k^\ast$ & $\alpha_1^\ast, \alpha_5^\ast$ & $\alpha_4^\ast$ & $\alpha_1^\ast$ & $\alpha_1^\ast$ & $\alpha_1^\ast$ \\
& & & & $\alpha^\ast_{n-1}$ & $\alpha_2^\ast, \alpha_4^\ast$ & $\alpha_6^\ast$ & & &  \\
& & &  & $\alpha^\ast_n$ & & $\alpha_7^\ast$ & &  &  \\
\hline
  \end{tabular}
 \caption{This table lists for every simple Lie algebra the charges $q$ associated to the coweights $x$. $q$ is defined by $c_2(B_x) = q H^2$ on $\mathbb{C}P^3$. The coweights that are not listed are associated to integer charges. The coweights are expressed in the basis dual to the root basis. The ordering of the roots is standard and follows Appendix 13.A of \cite{CFT1997}. The indices ranges are $n \geq 1$, $1 \leq k \leq n$ for $A_n$; $n \geq 3$, $1 \leq k \leq n$ for $B_n$; $n \geq 2$, $1 \leq k \leq n-1$ for $C_n$; $n \geq 4$, $1 \leq k \leq n-2$ for $D_n$.}
\label{t:QuantFracc2}
\end{table}
When comparing the table above to results pertaining to gauge theories on 4-manifolds \cite{Witten:2000nv}, one has to keep in mind that in the 4-dimensional context, the spacetime is generally assumed to be spin, while the cycle $\Sigma_4 \subset \mathbb{C}P^3$ dual to $H \cup H$ is not spin. The wedge product pairing of a 4-dimensional spin manifold is even and $c_2(K_{x,L})$ involves the square of the first Chern class of $L$, which generates an extra factor 2 compared to the table above. Note also that the bundle yielding half-instantons for $SO(8n)/\mathbb{Z}_2$ in \cite{Witten:2000nv} is not of the form $K_{x,L}$ which explains the absence of the factor 2 in this case.

\section{A relation modulo 12 in representation theory}

\label{AppRelMod12}

In this appendix, we restrict to a simple summand $\mathfrak{g}_i$ of the gauge Lie algebra and drop the subscript $i$: $b := b_i$, $\fg := \fg_i$, $\fR := \fR_i$. The gravitational anomaly coefficient $a$ and the anomaly coefficient $b$ satisfy:
\bea
a\cdot b &= {1 \ov 6} \left(-\sum_\fR x_\fR A_\fR + A_\text{Adj} \right) \\
b\cdot b &= {1 \ov 3} \left(\sum_\fR x_\fR C_\fR - C_\text{Adj} \right) \\
0 &= \sum_\fR x_\fR B_\fR - B_\text{Adj} \,.
\eea
where
\be
\label{DefABCApp}
{\tr_\fR F^2  = A_\fR \tr F^2}  \,, \quad
{\tr_\fR F^4 = B_\fR \tr F^4 + C_\fR (\tr F^2)^2}\;.
\ee
As before, $\tr$ is the trace in the adjoint representation of $\mathfrak{g}$ divided by twice the dual Coxeter number. $x_\fR$ is the multiplicity of the irreducible representation $\fR$ of $\mathfrak{g}$ in the matter spectrum. We actually have $A_\text{Adj} = B_\text{Adj} = 2h^\vee$, $C_\text{Adj} = 0$, where $h^\vee$ is the dual Coxeter number of $\mathfrak{g}$, but this will be irrelevant for what follows.

We now show that for some constant $\eta$ independent of $\fR$,
\be
\chi_\fR := 2 C_\fR - A_\fR + \eta B_\fR
\ee
is an integer divisible by 12 for any $\fR$, and thus
\be
a \cdot b + b\cdot b
= {\sum_\fR x_\fR \chi_\fR - \chi_\text{Adj} \ov 6}
\ee
is even.

Let $\theta$ be the highest root of $\mathfrak{g}$, which has length 2 by our normalization of the Killing form $(\bullet, \bullet)$. Let $H_\theta$ be the coroot corresponding to $\theta$. We have by definition
\be
\tr H_\theta^2 = (\theta, \theta) = 2 \;, \quad n_w := w(H_\theta) = (w, \theta) \in \mathbb{Z} \;,
\ee
for any integral weight $w$. $w(H_\theta)$ is the canonical pairing of the weight $w$ with the Cartan element $H_\theta$. We can rewrite \eqref{DefABCApp} as
\be
2A_\fR = \sum_{w \in \Lambda_\fR} d_w^\fR (w(H_\theta))^2 \;, \quad \sum_{w \in \Lambda_\fR} d_w^\fR (w(H_\theta))^4 = \frac{B_\fR}{2h^\vee} \sum_{v \in \Lambda_{\rm Adj}} d^{\rm Adj}_v (v(H_\theta))^4 + 4C_\fR
\ee
where the sums are over the set $\Lambda_\fR$ of weights of the representation $\fR$, $d_w^\fR$ is the multiplicity of $w$ in $\fR$.

Setting
\be
\eta = \frac{1}{4h^\vee}\sum_{v \in {\rm Adj}} d^{\rm Adj}_v (v(H_\theta))^4
\ee
we find that
\be
\chi_\fR = \frac{1}{2} \sum_{w \in \Lambda_\fR} d_w^\fR \left(  (w(H_\theta))^4 - (w(H_\theta))^2 \right) = \frac{1}{2} \sum_{w \in \Lambda_\fR} d_w^\fR (  n_w^4 - n_w^2 )
\ee
We can see $H_\theta$ is part of an $\mathfrak{sl}_2$ triple, so the weights $w$ of $\fR$ such that $n_w \neq 0$ come in pairs $(w_1, w_2)$ such that $n_{w_1} = - n_{w_2}$ and $d_{w_1}^\fR = d_{w_2}^\fR$. Writing $\Lambda^+_\fR$ for the weights $w$ of $\fR$ such that $n_w > 0$, we have
\be
\chi_R = \sum_{w \in \Lambda_\fR^+} d_w^\fR (  n_w^4 - n_w^2 )
\ee
As $n^4 - n^2$ is a multiple of 12 for all $n \in \mathbb{Z}$, we conclude that $\chi_\fR$ is a multiple of 12.

\section{$b_i \cdot b_{II} \in 2 \bZ$ from local anomalies} \label{ap:E}

We will show that for any representation $\mathfrak{R}_i$ of $\mathfrak{g}_i$ with $\mathfrak{g}_i \neq \mathfrak{sp}(n)$, $A^{\mathfrak{g}_i}_\mathfrak{R} = E^{\mathfrak{g}_i}_\mathfrak{R}$ modulo 2.  The anomaly equations
\be
\begin{aligned}
b_i \cdot b_{II} &= \sum_{\fR_i,q_I} x^{i,I}_{\fR_i,q_I}  q_I^2 A^{\mathfrak{g}_i}_{\fR_i} \,, \\
0&= \sum_{\fR_i, q_I} x^{i,I}_{\fR_i,q_I}  q_I  E^{\mathfrak{g}_i}_{\fR_i}
\end{aligned}
\ee
then immediately implies that $b_i \cdot b_{II} \in 2\mathbb{Z}$, which is Constraint \ref{Constr_8}. (Note that if $\mathfrak{g}$ has no third order Casimir, we trivially have $E_\mathfrak{R} = 0$ for all representations.)

Like in the previous appendix, we focus on a simple summand $\mathfrak{g}_i$ of the gauge Lie algebra and drop the subscript $i$. We assume that $\mathfrak{g}$ admits a $\mathfrak{su}(3)$ subalgebra of embedding index 1. This is essentially a $\mathfrak{su}(3)$ subalgebra whose roots coincides with long roots of $\mathfrak{g}$. In the following, it will be convenient to see $\mathfrak{su}(2)$ as $\mathfrak{sp}(1)$. Then for $\mathfrak{g}$ any simple Lie algebra different from $\mathfrak{sp}(n)$ or $\mathfrak{g}_2$, the corresponding Dynkin diagram admits $A_2$ as a subdiagram, which shows that such subalgebras exist. Moreover, the long roots of $\mathfrak{g}_2$ form a $\mathfrak{su}(3)$ subalgebra of the desired type. The argument below will therefore apply to any simple Lie algebra different from $\mathfrak{sp}(n)$.

Let us write $\fh$ for the said $\mathfrak{su}(3)$ subalgebra of $\fg$. Then any representation $\fR$ of $\fg$ is decomposed into representations $R$ of $\fh$:
\be
\fR = \oplus_R \, d^\fR_R R \,.
\ee
Thus, for an element $F \in \fh \subset \fg$, we get
\bea
\tr_\fR F^2 &=  \sum_R d^\fR_R \tr_R F^2
= \left( \sum_R d^\fR_R A^{\mathfrak{su}(3)}_R \right) \tr F^2 \,, \\
\tr_\fR F^3 &= \sum_R d^\fR_R \tr_R F^2
=\left(\sum_R d^\fR_R E^{\mathfrak{su}(3)}_R \right) \tr F^3 \,,
\eea
As the embedding index of $\mathfrak{h}$ is 1, the bilinear form $\tr$ coincides with its $\mathfrak{su}(3)$ counterpart when restricted on $\mathfrak{h}$. We can therefore write
\be\label{A from A}
A^\fg_\fR = {\tr_\fR F^2 \ov \tr F^2}
= \sum_R d^\fR_R A^{\mathfrak{su}(3)}_R \,,
\ee
\be
E^\fg_\fR = {\tr_\fR F^3 \ov \tr F^3} = \sum_R d^\fR_R E^{\mathfrak{su}(3)}_R \,.
\ee
It is therefore sufficient to prove that $A^{\mathfrak{su}(3)}_\mathfrak{R} = E^{\mathfrak{su}(3)}_\mathfrak{R}$ modulo 2 to show that  $A^\fg_\mathfrak{R} = E^\fg_\mathfrak{R}$ modulo 2 and prove Constraint \ref{Constr_8}.

We discard the superscript $\mathfrak{su}(3)$ for the rest of the section. As we are working with the algebra $\mathfrak{su}(3)$ we can be quite explicit. Let us represent the simple roots by the vectors
\be
\alpha_1 = \left({\sqrt{3} \ov \sqrt{2}}, {1 \ov \sqrt{2}}\right) \,, \quad
\alpha_2 = \left(-{\sqrt{3} \ov \sqrt{2}} , {1 \ov \sqrt{2}}\right) \,.
\ee
The fundamental weights are then
\be
e_1 = \left( {1 \ov \sqrt{6}}, {1 \ov\sqrt{2}}\right) \,, \quad
e_2 = \left( -{1 \ov \sqrt{6}}, {1 \ov \sqrt{2}}\right) \,.
\ee
The weights of any representation $R$ of $\mathfrak{su}(3)$ are given by $w_{n_1,n_2} = n_1 e_1 + n_2 e_2$, for integers $\{n_i\}$ and have multiplicity $d^R_{n_1,n_2}$. In the fundamental representation, the weights are given by $w_{1,0}$, $w_{0,-1}$ and $w_{-1,1}$, each with multiplicity 1.

Then, by taking $F = (1,0)$ in the Cartan subalgebra of $\mathfrak{su}(3)$, we find that
\bea
A_R &= {{\sum_{w_{n_1,n_2} \in \Lambda_R} d^R_{n_1,n_2} \left({n_1 - n_2 \ov \sqrt{6}}\right)^2}
\ov
{\sum_{w_{n_1,n_2} \in \Lambda_\mathbf{3}} d^\mathbf{3}_{n_1,n_2} \left({n_1 - n_2 \ov \sqrt{6}}\right)^2} }
= {1 \ov 6} \sum_{w_{n_1,n_2} \in \Lambda_R} d^R_{n_1,n_2} \left({n_1 - n_2 }\right)^2 \,, \\
E_R &= {{\sum_{w_{n_1,n_2} \in \Lambda_R} d^R_{n_1,n_2} \left({n_1 - n_2 \ov \sqrt{6}}\right)^3}
\ov
{\sum_{w_{n_1,n_2} \in \Lambda_\mathbf{3}} d^\mathbf{3}_{n_1,n_2} \left({n_1 - n_2 \ov \sqrt{6}}\right)^3} }
= -{1 \ov 6} \sum_{w_{n_1,n_2} \in \Lambda_R} d^R_{n_1,n_2} \left({n_1 - n_2 }\right)^3 \,.
\eea

The weights in $\Lambda_R$ decompose into orbits of the Weyl group, the generators of which takes $w_{n_1,n_2}$ to $w_{n_1+n_2,-n_2}$ and $w_{-n_1,n_1+n_2}$, respectively. Note that two weights related by the Weyl symmetry have the same degeneracy. There are three types of orbits of the Weyl group. The first is the zero orbit, where $n_1 = n_2 =0$. It has one element. There are also orbits of length $3$, which form equilateral triangles in the plane. These orbits are labeled by a single integer $n$ and the corresponding weights are given by
\be
w_{n,0},~
w_{0,-n},~
w_{-n,n} \,.
\ee
We write $\cO^R_3$ for the set of orbits of length 3 in the representation $R$. Finally, there are orbits of length $6$, labeled by a pair of integers $n_1 > n_2 > 0$. The weights in the orbit are given by
\be
w_{n_1,n_2},~
w_{n_1 + n_2,-n_2},~
w_{-n_1,n_1 + n_2},~
w_{n_2,-n_1-n_2},~
w_{-n_2,-n_1},~
w_{-n_1-n_2,n_1} \,.
\ee
We write $\cO^{R}_6$ for the set of orbits of length 6 in the representation $R$. We can compute
\bea
A_R &= {1 \ov 6} \sum_{w_{n_1,n_2} \in \Lambda_R} d^R_{n_1,n_2} \left({n_1 - n_2 }\right)^2 \\
&= {1 \ov 6} \sum_{n \in \cO^R_3} d^R_{n,0}  \left(  n^2 + n^2+(-2n)^2 \right) \\
&\quad + {1 \ov 6} \sum_{(n_1,n_2) \in \cO^R_6} d^R_{n_1,n_2}
\left(  2(n_1-n_2)^2 + 2(-2n_1-n_2)^2+2(2n_2+n_1)^2 \right) \\
&= \sum_{n \in \cO^R_3} d^R_{n,0} \cdot  n^2 +
\sum_{(n_1,n_2) \in \cO^R_6} d^R_{n_1,n_2}
\cdot 2 \left( n_1^2 +n_2^2 + n_1 n_2\right)
\equiv \sum_{n \in \cO^R_3} d^R_{n,0} \cdot  n^2 ~\text{(mod 2)} \,.
\eea
Likewise, we find that
\bea
E_R
&= -\sum_{n \in \cO^R_3} d^R_{n,0} \cdot  n^3 +
\sum_{(n_1,n_2) \in \cO^R_6} d^R_{n_1,n_2}
\cdot \left( -2n_1^3 -3n_1^2 n_2 + 3 n_1 n_2^2 + 2 n_2^3\right)  \\
&\equiv -\sum_{n \in \cO^R_3} d^R_{n,0} \cdot  n^3 ~\text{(mod 2)} \,,
\eea
where we have used the fact that $n_1 n_2 (n_1 - n_2)$ is always even. We thus find that $E_R$ is an integer. Furthermore, since $d^R_{n,0}$ are integers and $n^3$ and $n^2$ have the same parity,
\be
A_R \equiv E_R ~\text{(mod 2)} \,,
\ee
as desired.

\bibliographystyle{JHEP}
\bibliography{refs}

\end{document}